\documentclass[12pt]{iopart}
\usepackage{iopams}
\usepackage{graphicx}
\usepackage{mathrsfs}
\usepackage{amssymb}
\usepackage{stackrel}
\usepackage{cite}
\usepackage[section]{placeins}
\usepackage{xcolor}
 \newcommand{\vek}[1]{\mbox{\boldmath$#1$}}
\newcommand{\textchg}[2]{#2}
\begin{document}
\title{Bayesian inference \textchg{red}{of scaled versus fractional} Brownian motion}
\author{Samudrajit Thapa$^{1,2}$, Seongyu Park$^3$, 
Yeongjin Kim$^3$, Jae-Hyung Jeon$^3$, Ralf Metzler$^4$,
Michael A. Lomholt$^5$\footnote{E-mail: mlomholt@sdu.dk}}
\address{$^1$ Sackler Center for Computational Molecular and Materials Science, Tel Aviv University, Tel Aviv 6997801, Israel}
\address{$^2$ School of Mechanical Engineering, Tel Aviv University, Tel Aviv 6997801, Israel}
\address{$^3$ Department of Physics, Pohang University of Science and Technology (POSTECH), Pohang 37673, Republic of Korea}
\address{$^4$ Institute of Physics and Astronomy, University of Potsdam,
D-14476 Potsdam-Golm, Germany}
\address{$^5$ PhyLife, Department of Physics, Chemistry and Pharmacy,
University of Southern Denmark, Campusvej 55, 5230 Odense M, Denmark}

\begin{abstract}
We present a Bayesian inference scheme for scaled Brownian motion, and investigate its performance on synthetic data for parameter estimation and model selection in a combined inference with fractional Brownian motion. We include the possibility of measurement noise in both models. 
We find that for trajectories of a few hundred time points the procedure is able to resolve well the true model and parameters. Using the prior of the synthetic data generation process also for the inference, the approach is optimal based on decision theory. We include a comparison with inference using a prior different from the data generating one.
\end{abstract}

\section{Introduction}
With Robert Brown’s observation of the random motion of micron-sized granules contained in pollen grains \cite{brown1828}, Albert Einstein’s explanation of the physical origin of Brownian motion \cite{einstein05}, and Jean Perrin’s \cite{perrin09} and Ivar Nordlund's \cite{nordlund14} subsequent quantitative measurements, the field of single particle tracking was born. More recently, with the advent of fluorescence microscopy, it has become possible to track single particles inside highly complex environments such as biological cells. This has revealed motion distinct from pure Brownian, and a wealth of anomalous diffusion models have been put forward to quantify this behaviour \cite{metzler2014,sokolov12}. The question then arises, how does one determine which of these models best fit the observed random motion?

Mathematical models of randomly fluctuating data---such as the position time series measured in single particle tracking experiments---are comprehensibly described by writing down the probability of the complete data given the model: $P({\rm data}|{\rm model})$. But when we do inference we would like to make deductions in the opposite direction. In Bayesian inference such deductions are quantified by allowing probability to be interpreted as belief in different models (including the parameter values necessary to completely specify the model), and then the beliefs are updated when data is revealed through the use of Bayes' formula, which up to a normalisation constant says that $P({\rm model}|{\rm data})\propto P({\rm data}|{\rm model})P({\rm model})$ \cite{mackay03,sivia06,robert07,gelman13}. The main controversies around Bayesian inference originates from the need to assign probabilities $P({\rm model})$ that represent prior beliefs in the model in the absence of any data. For this assignment, there is no universally accepted procedure.

The objections leading to this controversy, however, are not valid when the data are generated by procedures where probability distributions for all random assignments are known. In this case $P({\rm model})$ is a known probability distribution and Bayes' formula has the status of a mathematical theorem with satisfied prerequisites. The problem of inferring a parameter value in a way that optimises the average value of a certain score function thus becomes a well-defined problem in decision theory with an optimal solution \cite{robert07,elliott16}.

In the AnDi Challenge \cite{andi21} artificial trajectories are generated from five disclosed models of anomalous diffusion with announced limits on the model parameters. The precise parameter distributions, however, are not disclosed. This introduces some psychological guesswork into the competition, since arbitrary choices have to be made, for instance, for the Bayesian parameter priors. Similarly, for machine learning methods choices also have to be made for the distributions of parameters with which the training sets are generated. However, the AnDi Challenge is very close to having disclosed the data generating procedure completely. Thus one would expect that an approach based on Bayesian inference and decision theory in principle should be able to come very close to an optimal solution.

Although an approach based on Bayesian inference and decision theory is optimal in principle, carrying out the necessary computations to evaluate the solution can be a difficult problem, in particular when the model contains hidden variables that needs to be integrated out. Thus the AnDi Challenge team that many of the present authors participated in (BIT) did not succeed in implementing effective computational solutions for all five models in time for the challenge.

\begin{figure}
   \centering
    \includegraphics[scale=0.5]{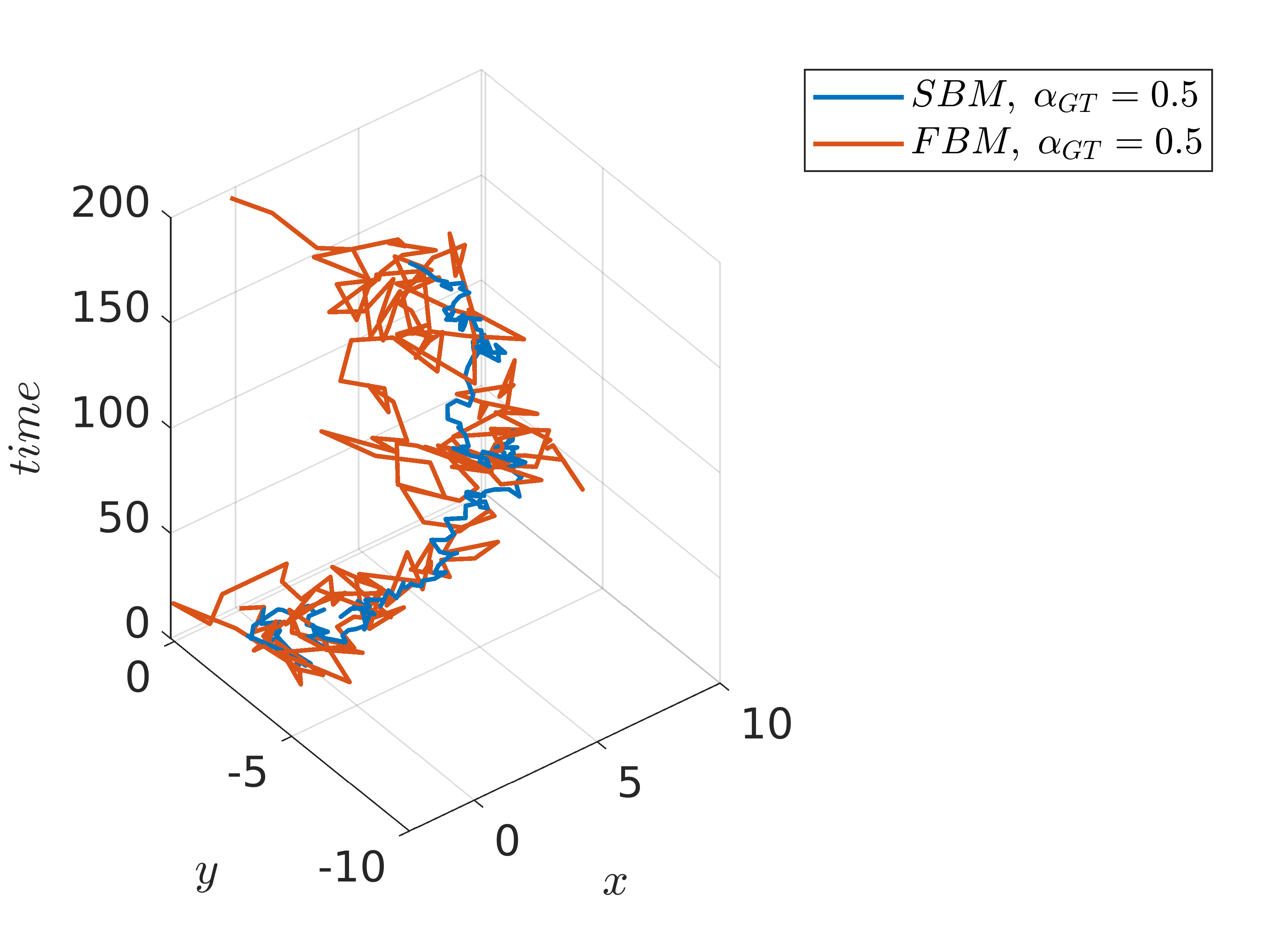}
    \includegraphics[scale=0.5]{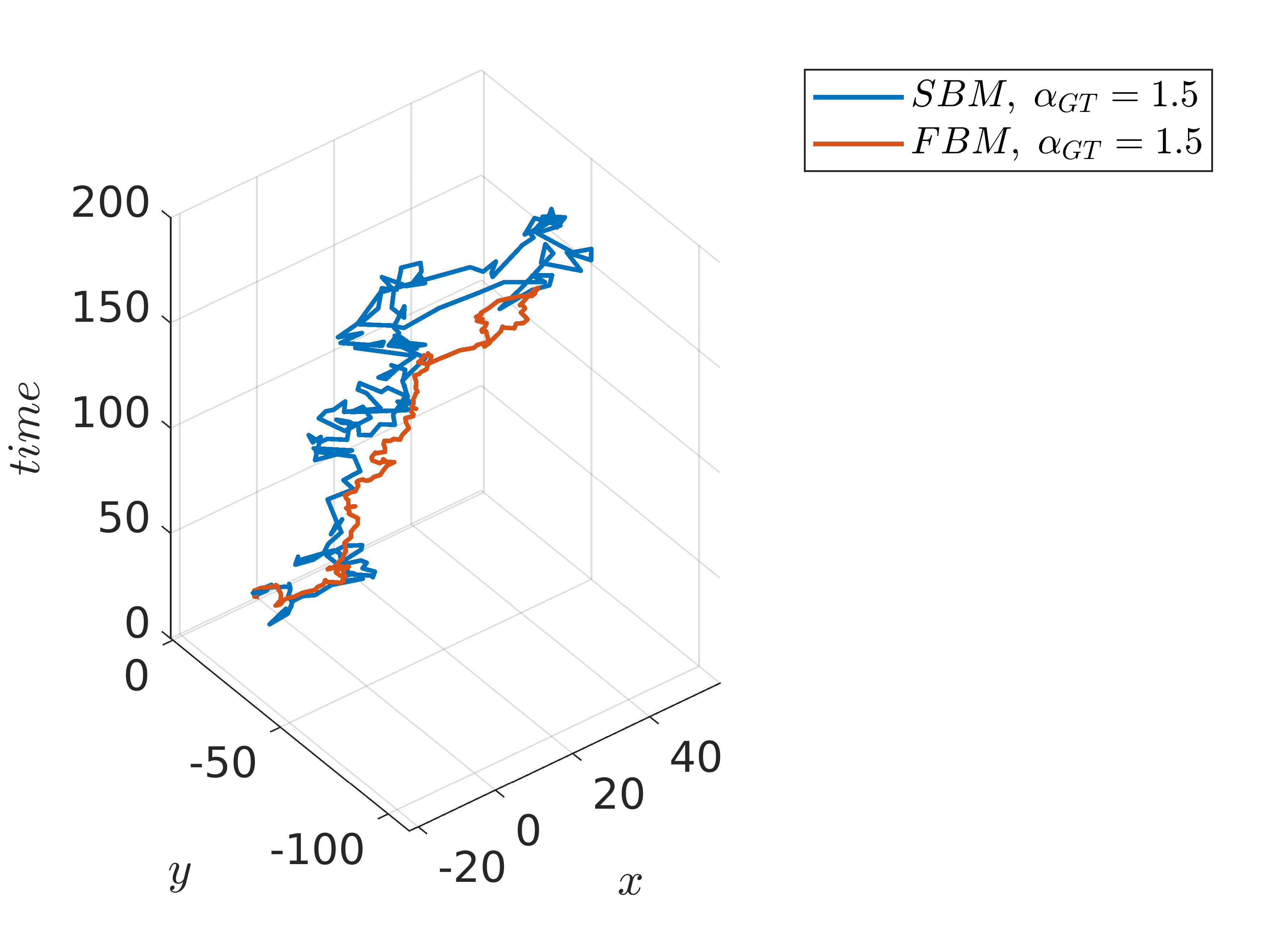}
    \caption{ Sample trajectories for subdiffusive (left) and superdiffusive (right) 2-dimensional SBM and FBM  with noise strength $\sigma_{mn}=0.1$. We show trajectories of FBM and SBM that look quite similar.
    The trajectories were chosen from 4 independent realisations for the subdiffusive case and from 5 independent realisations for the superdiffusive case. Each trajectory consists of $N=200$ points. The SBM trajectories were generated with $t_0=0$ and $\sigma_1=1$ while the FBM trajectories were generated with $\sigma=1$. See sections \ref{sec-sbm}, \ref{sec-sbm-mn} and \ref{sec-fbm} for details on the parameters.}
    \label{fig-trajs-2d}
\end{figure}

In this article we present our Bayesian inference approach for one of the AnDi challenge models: scaled Brownian motion (SBM)\footnote{We refer interested readers to  \cite{krog17,krog18,thapa2018,park2021} for details of our implementation of Bayesian inference for other models of diffusion.}. SBM is a simple Markovian model of anomalous diffusion, where Brownian motion is modified by allowing the diffusion coefficient to depend on time \cite{lim02}. This approach with time dependent diffusion coefficient is a convenient way to generalise for instance formulas for photobleaching recovery data \cite{saxton01}. We combine our implementation for SBM with the implementation of Bayesian inference for fractional Brownian motion (FBM) presented in \cite{krog18} to demonstrate model selection and inference of the anomalous diffusion exponent for these two models. FBM is a modification of Brownian motion, where correlations are allowed for the motion at different times \cite{mandelbrot68}. Figure \ref{fig-trajs-2d} shows examples for  subdiffusive and superdiffusive 2-dimensional trajectories of SBM and FBM. The trajectories of these processes can look qualitatively different because while SBM is a process with non-stationary increments, FBM has stationary increments. Subdiffusive SBM diffuses progressively slowly with time, whereas superdiffusive SBM diffuses progressively fast. However, the 
trajectories of FBM and SBM can look similar, as is shown in figure \ref{fig-trajs-2d} by selecting trajectories that look alike from a few realisations.

The article is organised with our inference methods being presented in Section \ref{sec:methods}, before we present results showing how effective they are for parameter estimation and model selection in Section \ref{sec:results} and conclude in Section \ref{sec:concl}.

\section{Methods}\label{sec:methods}
In this section we briefly review Bayesian inference before presenting the likelihood functions of the models in subsections \ref{sec-sbm}--\ref{sec:multiple}. Finally, we state our computational methods.

\subsection{Bayesian inference}
If we have a time series $\Delta\vek{x}_N$ of $N$ steps  generated from a known model $M_i$ with parameters $\vek{\theta}$ taken from a known probability distribution $\pi_i(\vek{\theta})=P(\vek{\theta}|M_i)$, Bayes' theorem tells us that the probability distribution (called posterior) for the unknown parameter $\vek{\theta}$ is
\begin{equation}
\label{eq-posterior}
P(\vek{\theta}|\Delta\vek{x}_N,M_i)=\frac{\mathcal{L}_i(\vek{\theta})\pi_i(\vek{\theta})}{E_i},
\end{equation}
where $\mathcal{L}_i(\vek{\theta})=P(\Delta\vek{x}_N|\vek{\theta},M_i)$ is called the likelihood function and the normalisation constant (called evidence) is $E_i=P(\Delta\vek{x}_N|M_i)=\int d\vek{\theta}\,\mathcal{L}_i(\vek{\theta})\pi_i(\vek{\theta})$. If, additionally, the model $M_i$ is randomly chosen from a set of $M$ models: $M_i$, $i=1,\dots,M$, with probabilities $\pi(M_i)$, then we can again use Bayes' theorem to find the probability $P(M_i|\Delta\vek{x}_N)$ that the model was $M_i$ given an artificially generated trajectory $\Delta\vek{x}_N$,
\begin{equation}
\label{eq-model-prob}
P(M_i|\Delta\vek{x}_N)=\frac{E_i \pi(M_i)}{\sum_{i=1}^M E_i \pi(M_i)}.
\end{equation}
If the priors $\pi_i(\vek{\theta})$ and $\pi(M_i)$ are not known, for instance, if the trajectory is experimental data from a system that is not well known, then Bayesian inference requires that we interpret the priors as quantifying our lack of knowledge about the parameters and the true model. For instance, if we have no basis for assuming that one model is more probable than another, we should choose $\pi(M_i)=1/M$. For a continuous parameter with unlimited range the choice of a uniform prior is generally not possible. In this case one then has to shape the prior non-uniformly according to judgement about realistic values of the parameter.

\subsection{Scaled Brownian motion}
\label{sec-sbm}
Scaled Brownian motion (SBM) is a popular stochastic process that has been used to model anomalous diffusion 
observed in experiments where the diffusion coefficient seems to be time-dependent \cite{weiss07,verkman98,berland08,hoyst06,doussal92,boon13}. It is particularly convenient to describe experiments with fluorescence recovery after photobleaching \cite{saxton01} \textchg{red}{in which the diffusion coefficient appears to depend on time as a power law: $D(t)=\alpha K_\alpha(t+t_0)^{\alpha-1}$. Here $K_\alpha$ is the anomalous diffusion coefficient, $\alpha$ the anomalous diffusion exponent, and $t_0$ the ageing time of the system prior to $t=0$ (all taken to be real and positive numbers).
Scaled Brownian motion in one dimension can be described mathematically as having the} mean squared displacement
\begin{equation}
\langle [x(t)-x(0)]^2\rangle = 2 K_\alpha [(t+t_0)^\alpha-t_0^\alpha]\label{eq:msd}
\end{equation}
and zero mean independent Gaussian increments \cite{lim02}. \textchg{red}{The corresponding probability density $P(x,t)$ for the position $x$ satisfies a diffusion equation with explicit time dependence
\begin{equation}
    \frac{\partial P}{\partial t}=D(t)\frac{\partial^2 P}{\partial x^2}\label{eq:diff}
\end{equation}
and t}he increments are not stationary in general since
\begin{equation}
\langle [x(t+\Delta t)-x(t)]^2\rangle = 2 K_\alpha [(t+\Delta t+t_0)^\alpha-(t+t_0)^\alpha]\label{eq:inc_var}
\end{equation}
depends on the time $t$ unless we are in the case of pure Brownian motion with $\alpha=1$. SBM exhibits a number of interesting properties, such as ageing and non-ergodic behaviour. 
Regarding ageing, this can be seen from Eq. (\ref{eq:inc_var}), since the statistics of the increments depend on the time $t+t_0$ since initiation of the system, which is the characteristic feature of ageing. For a further discussion of these properties we refer the reader to \textchg{red}{\cite{hyung14,safdari2015,cherstvy_sbm2015,bodrova2015}}.

Since the increments are independent \textchg{red}{and Gaussian}, we can obtain the probability density for a set of increments $\Delta\vek{x}_N=(\Delta x_1,\dots,\Delta x_N)$ where $\Delta x_i=x_i-x_{i-1}$ and $x_i=x(t_i)$ for equidistant times $t_i=i \Delta t$. The \textchg{red}{probability density is a product of the densities for each increment,}
\begin{equation}
P(\Delta\vek{x}_N|\vek{\theta},M_1)=\prod_{i=1}^N\frac{1}{\sqrt{2\pi \sigma_i^2}}\mathrm{exp}\left(-\frac{\Delta x_i^2}{2{\sigma}_i^2} \right),
\end{equation}
where we have labelled the SBM model $M_1$, $\vek{\theta}=(\alpha,K_\alpha,t_0)$ and $\sigma_i^2=2 K_\alpha \Delta t^\alpha [(i+t_0/\Delta t)^\alpha-(i-1+t_0/\Delta t)^\alpha]$.

\subsection{Scaled Brownian motion with measurement noise}
\label{sec-sbm-mn}
\textchg{red}{The experimental observation of single particle motion is complicated by noise in the measurement procedure \cite{michalet10,michalet12}. 
Here we} model measurement noise as independent Gaussian displacements $\eta_i$ of the underlying true positions $x_i^{\mathrm{true}}$ resulting in a measured position at time $t_i$ which is
\begin{equation}
    x_i = x_i^{\mathrm{true}} + \eta_i,
\end{equation}
where $x_i^{\mathrm{true}}$ follows the statistics of SBM as presented in the previous subsection and $\eta_i$ are independent with zero mean and variance $\langle \eta_i^2\rangle=\sigma_{\rm mn}^2$. For this situation, we can find a formula for the probability of the observed trajectory
\begin{equation}
P(\Delta\vek{x}_N|\vek{\theta},M_1)=\prod_{i=1}^N \frac{1}{\sqrt{2\pi \tilde{\sigma}_i^2}}\mathrm{ exp}\left(-\frac{1}{2\tilde{\sigma}_i^2}(\Delta x_i - \Delta \tilde{x}_i)^2 \right),
\end{equation}
where we have recursively for $i\ge 1$
\begin{eqnarray}
\Delta \tilde{x}_{i+1} &=& \mathrm{E}(\Delta x_{i+1}|\Delta \vek{x}_{i}) = - \frac{\sigma_{\rm mn}^2}{\tilde{\sigma}_{i}^2}(\Delta x_{i} - \Delta \tilde{x}_{i})\,, \label{eq:meanx}\\
\tilde{\sigma}_{i+1}^2 &=& \mathrm{Var}(\textchg{red}{\Delta} x_{i+1}|\Delta \vek{x}_{i}) =  \sigma_i^{2} + \sigma_{\rm mn}^2\left(2-\frac{\sigma_{\rm mn}^2}{\tilde{\sigma}_{i}^2} \right),\label{eq:varx}
\end{eqnarray}
with base case $\tilde{\sigma}_{1}^2={\sigma}_{1}^2+2 {\sigma}_{\rm mn}^2$ and $\Delta \tilde{x}_{1}=0$. Here $\mathrm{E}(a|b)$ means \textchg{red}{the expected value} of $a$ given $b$ and $\mathrm{Var}(a|b)$ means variance of $a$ given $b$. The derivation of the above formulas is a straightforward generalisation of the derivation in \cite{krog17}, where the ${\sigma}_{i}$ are assumed to be identical. 
This is the case since according to the derivation in \cite{krog17} the joint probability distribution for $x_{i+1}$ and $\eta_{i+1}$ at time $t_{i+1}$ only depends on the corresponding distribution at time $t_i$ and what happens during the time interval to $t_{i+1}$.

Note from Eq. (\ref{eq:meanx}) that the measurement noise introduces anti-correlations between steps: if a step was further to the right than expected, then the next step will tend to be to the left. This is similar to subdiffusive fractional Brownian motion (i.e., with $\alpha<1$), but there the correlations are long-range.

\subsection{Fractional Brownian motion}
\label{sec-fbm}
{Fractional Brownian motion (FBM) \cite{mandelbrot68,mandelbrot82} is a generalisation of Brownian motion to include correlated increments that can successfully  model anomalous diffusion in numerous experiments, particularly in visco-elastic systems \cite{cherstvy2019,jeon2011,weiss09,weron09,klafter10,burn10,weber10,korabel20,weiss-ent-21,weiss-njp-21}.}
FBM with measurement noise is a stationary Gaussian process whose increments are correlated with covariance $\langle \Delta x_i \Delta x_{j}\rangle=\gamma(i-j)$ where
\begin{equation}
\gamma(n) = \left\{\begin{array}{l l} 2 K_\alpha \Delta t^\alpha+2\sigma_{\rm mn}^2, & n=0\\
K_\alpha \Delta t^\alpha(2^\alpha-2)-\sigma_{\rm mn}^2, & |n|=1\\
K_\alpha \Delta t^\alpha (|n+1|^\alpha+|n-1|^\alpha-2|n|^\alpha), & |n|\ge 2
\end{array}\right.
\end{equation}
with $K_\alpha$ being the diffusion coefficient \cite{krog18}. Note that for $\alpha=1$, this becomes identical to SBM with $\alpha=1$ and arbitrary $t_0$. The probability for a trajectory (the likelihood function) can be written as
\begin{equation}
P(\Delta\vek{x}_N|\vek{\theta},M_2)=\frac{1}{(2\pi)^{N/2}|\vek{\Gamma}_N|^{1/2}}\exp\left(-\frac{1}{2}\Delta\vek{x}_N^T\vek{\Gamma}_N^{-1}\Delta\vek{x}_N\right),
\end{equation}
where $\vek{\Gamma}_N$ is the covariance matrix with components $\Gamma_{N,ij}=\gamma(i-j)$ and we have labelled FBM as $M_2$. FBM does not display ageing, and it is ergodic, albeit with slow convergence of the time-averaged mean-squared displacement to its ensemble averaged limit \cite{deng09,jeon2012,jeon2013}. \textchg{red}{Despite these differences with SBM it obeys the same time dependent diffusion equation, Eq. (\ref{eq:diff}), and has mean square displacement as in Eq. (\ref{eq:msd}), but with no effect of ageing, i.e., $t_0=0$.} To calculate the likelihood function \textcolor{red}{for FBM} numerically, we employ the Durbin-Levinson algorithm as done previously for Bayesian inference with FBM in \cite{krog18}.

\subsection{Multiple dimensions}
\label{sec:multiple}

The likelihood functions for SBM and FBM described above straightforwardly generalises to multiple dimensions, by assuming independent steps in each dimension. The likelihood function then becomes a product of the likelihood functions in each dimension with shared $\alpha$, $K_\alpha$ and possibly $t_0$. For instance in two dimensions we will have
\begin{equation}
\mathcal{L}_i(\vek{\theta}) = P(\Delta\vek{x}_N|\vek{\theta},M_i) P(\Delta\vek{y}_N|\vek{\theta},M_i)
\end{equation}
where $\Delta \vek{y}_N$ contains the coordinates along the second dimension.
Keeping in mind the large number of single particle tracking experiments in two dimensions \cite{manzo2015, lene2017}, we will mainly consider two dimensions in the following. The results are similar in other dimensions and we present results from 1-dimensional trajectories in the appendix.

\subsection{Computational methods}
To evaluate the model evidences $E_i$ and sample from the posteriors $P(\vek{\theta}|\Delta \vek{x}_N,M_i)$ we use the nested sampling algorithm of Skilling \cite{skilling04,skilling06}. Our implementation is an update of the one presented in \cite{krog18}, where the random walk Monte Carlo steps have been replaced by the discontinuous Hamiltonian Monte Carlo method presented in \cite{nishimura20}. We have uploaded our implementation to GitHub \cite{sbm_git}.

\section{Results}\label{sec:results}
Here we specify our priors and parameters for the synthetic data generation before giving our results for the parameter estimation in subsection \ref {sec-param-estim} and model selection in \ref{subsec:model}. Finally, we test our implementation on continuous time random walk trajectories, which illustrates the use of Bayesian inference on data not generated by a model included in the inference.

\subsection{Priors}
\label{sec-priors}
We choose a uniform prior on $\alpha$ restricted to  $0<\alpha<2$ for both SBM and FBM. We parametrise the diffusion coefficient $K_\alpha$  of FBM as $K_\alpha = \sigma^2/(2\Delta t^\alpha)$ with $\log_{10}\sigma$ having a standard normal distribution, where $\sigma$ is the step-deviation. In the case of SBM, the step-deviation $\sigma_i$ (and therefore diffusion coefficient) changes with time. We choose the standard deviation of the first step $\sigma_1$ as a parameter with $\log_{10}\sigma_1$ having a standard normal distribution.  For the noise strength $\sigma_{\rm mn}$ we choose a uniform prior in the range $0<\sigma_{\rm mn}<1$ for the analysis of all simulated data sets except for those generated with
$\sigma_{\rm mn}=10$. We choose a uniform prior on $\sigma_{\rm mn}$ in the range $0<\sigma_{\rm mn}<10$ for the latter case, such that
the true value of the noise-strength lies within the prior range. For the analysis with a wrong prior on $\alpha$ (see below) we use a linear prior $p(\alpha)=\frac{\alpha}{2}$ with $0 < \alpha <2$. Finally, the models have prior $\pi(M_i)=1/2$.

\subsection{Details of simulated data sets}
\label{sec-datasets}
In order to quantify the performance of our code, we simulated SBM and FBM trajectories to constitute data sets that were used in the analysis described in the subsequent sections.  All the SBM trajectories were generated with $t_0=0$, $\Delta t =1$ and $\log_{10}\sigma_1$ drawn from a standard normal distribution. The FBM trajectories were also generated with $\Delta t =1$ and $\log_{10}\sigma$ drawn from a standard normal distribution.  The other simulation parameters such as $\alpha_{\rm GT}$, $N$ and $\sigma_{\rm mn}$ are specified in the subsequent sections for each data set
used in the analysis. \textchg{red}{We remark here that a choice of $\sigma_{\rm mn}=0$ means no measurement noise, while for $\sigma_{\rm mn}=1$ the measurement noise will typically be comparable in strength to $\sigma_1$, i.e., the diffusive noise of the first single step.} The codes for the simulations of both FBM and SBM trajectories are home-written and presented in \cite{sbm_git}. The FBM generation algorithm is the same that was used in \cite{krog18}.

\subsection{Quantification of parameter estimation}
\label{sec-param-estim}
To quantify the parameter estimation results we use the mean absolute error (MAE), which for a parameter $\theta$ is defined as \cite{andi21}
\begin{equation}
\label{eq-mae}
    {\rm MAE}=\frac{1}{\tilde{N}}\sum_{j=1}^{\tilde{N}}\left|\theta_{j,p}-\theta_{j,{\rm GT}} \right|,
\end{equation}
where the subscript $j$ denotes the trajectory number, $\theta_{j,p}$ is the inferred 
parameter value, $\theta_{j,{\rm GT}}$ is the true value (ground truth) and $\tilde{N}$ is the
total number of trajectories. In what follows we choose $\tilde{N}=100$ trajectories for each MAE estimate,
except when we compare the results in the case of different priors on $\alpha$.  For the latter case we choose $\tilde{N}=1000$ for each MAE estimate. This is because, to highlight the effect of the choice of prior distribution, we need a sufficiently large number of trajectories to achieve a significant difference between the two priors. 

In order to compute the MAE in Eq. (\ref{eq-mae}) we consider the median of the posterior distribution (Eq. (\ref{eq-posterior})) weighted with the corresponding model probability (Eq. (\ref{eq-model-prob})) as the inferred value. Therefore the inferred $\alpha$ from a particular trajectory $j$ is obtained as
\begin{equation}
\label{eq-alpha-inference}
    \alpha_{j,p}= \tilde{\alpha}_{ j,1}P(M_1|(\Delta\vek{x}_N)_j)+\tilde{\alpha}_{ j,2}P(M_2|(\Delta\vek{x}_N)_j),
\end{equation}
where $\tilde{\alpha}_{j,i}$ is the median of the posterior distribution of $\alpha$ for model $M_i$ and trajectory $j$, whose steps are collected in $(\Delta\vek{x}_N)_j$. We have chosen the median of the posterior as estimator of $\alpha$, since this choice is the optimal one based on decision theory, i.e., on average it minimises the MAE \cite{robert07,elliott16}.

\begin{figure}
    \includegraphics[angle=270,scale=0.35]{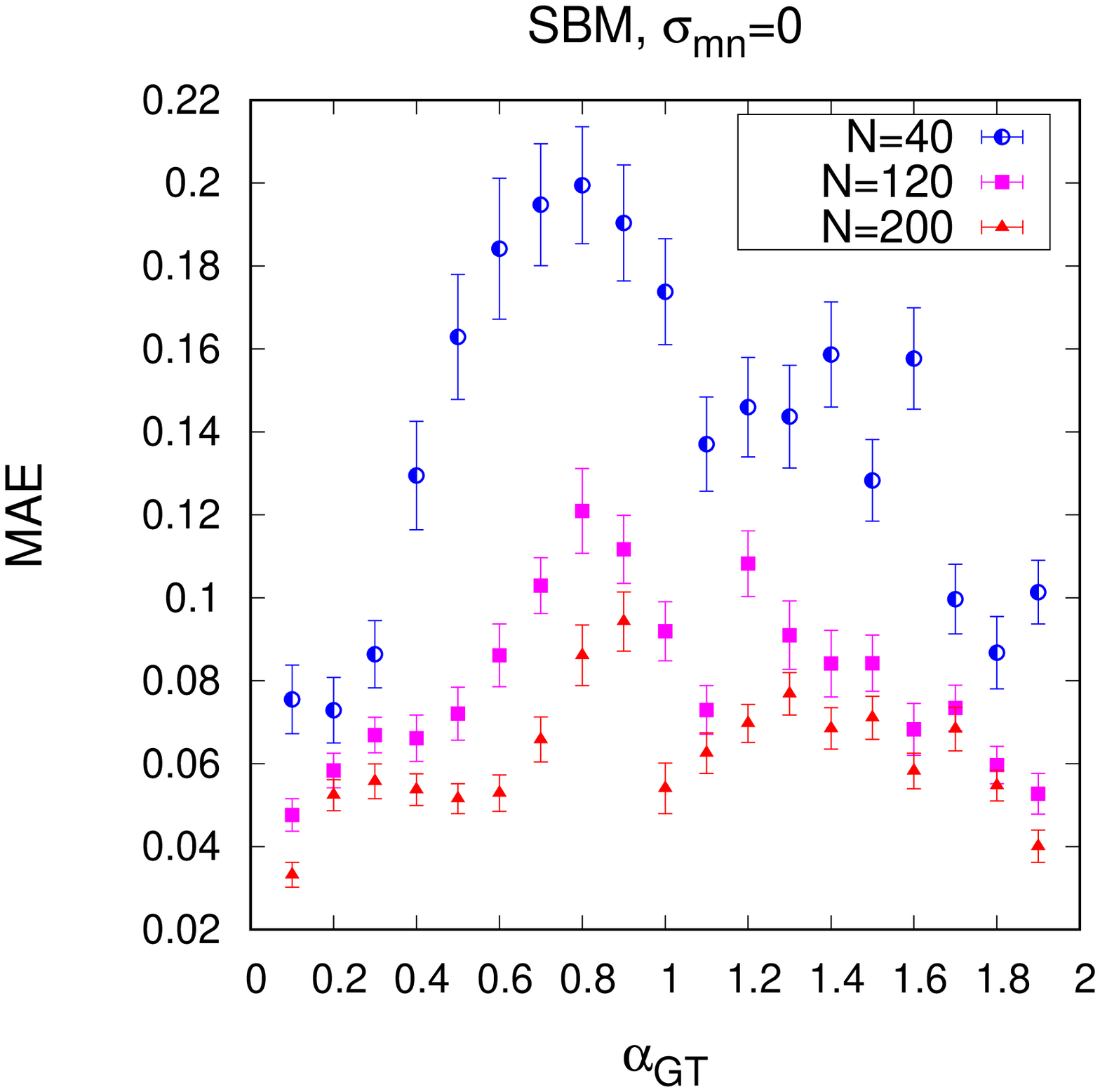} 
    \includegraphics[angle=270,scale=0.35]{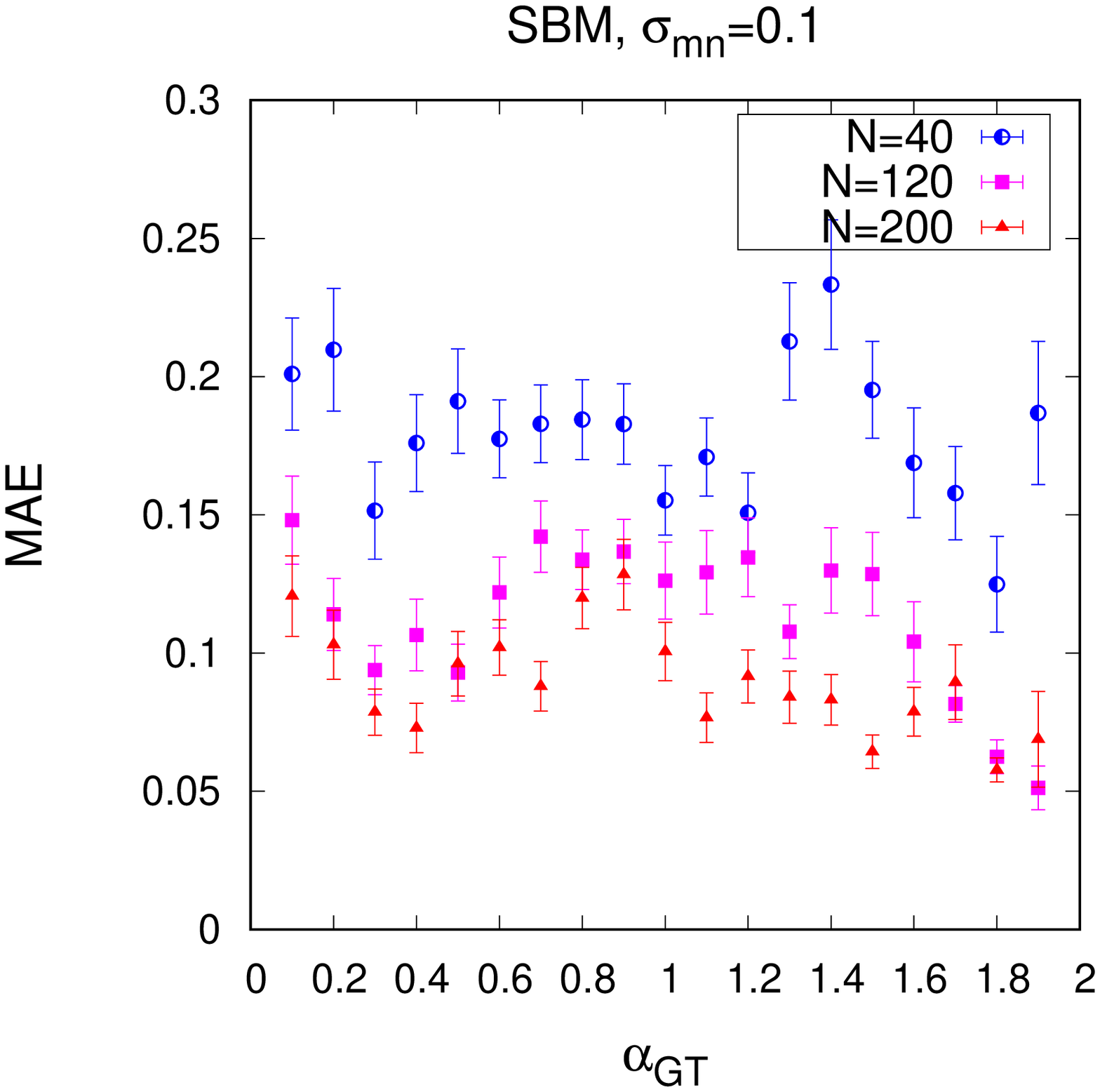}\\ 
    \includegraphics[angle=270,scale=0.35]{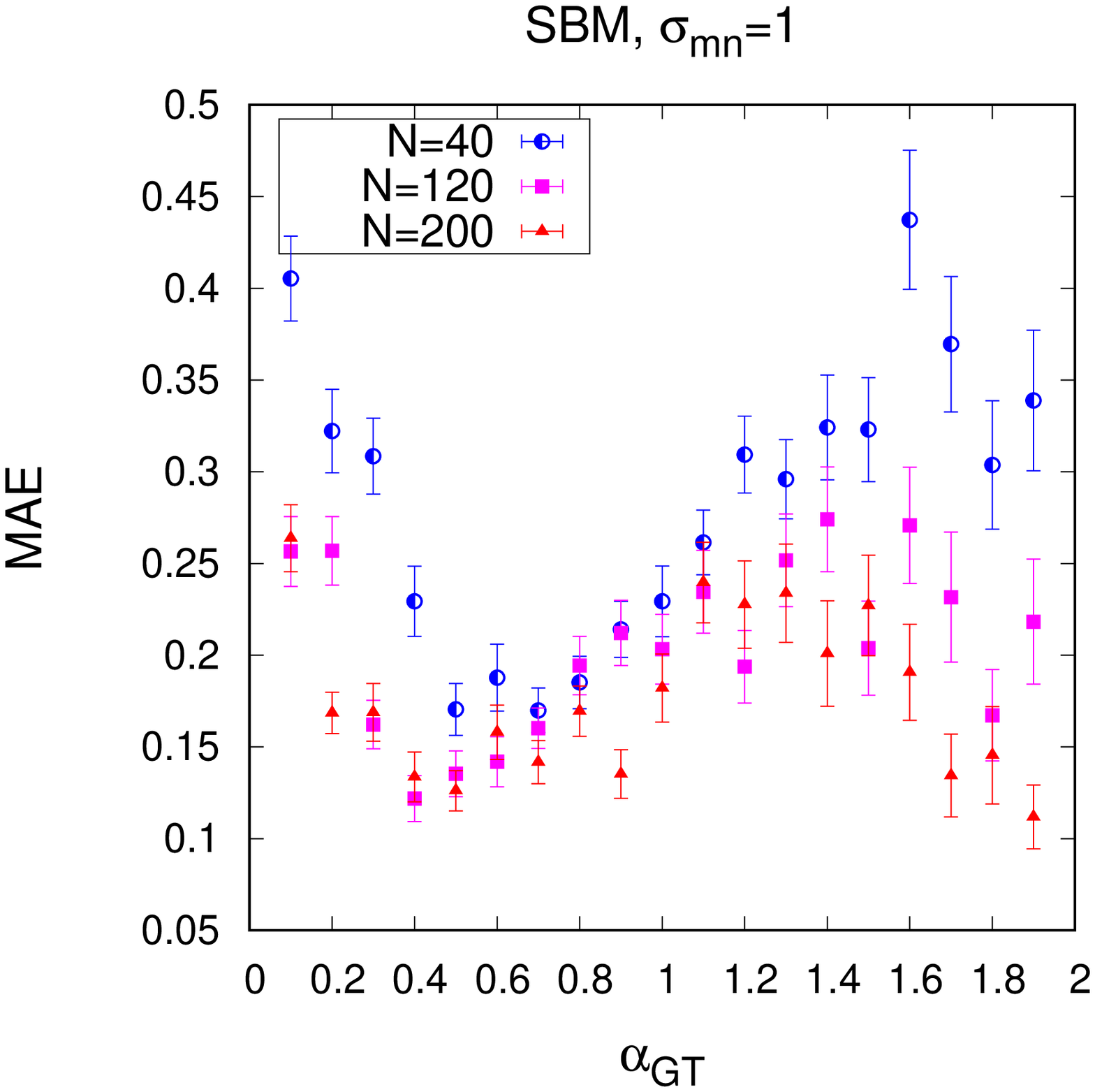} 
    \includegraphics[angle=270,scale=0.35]{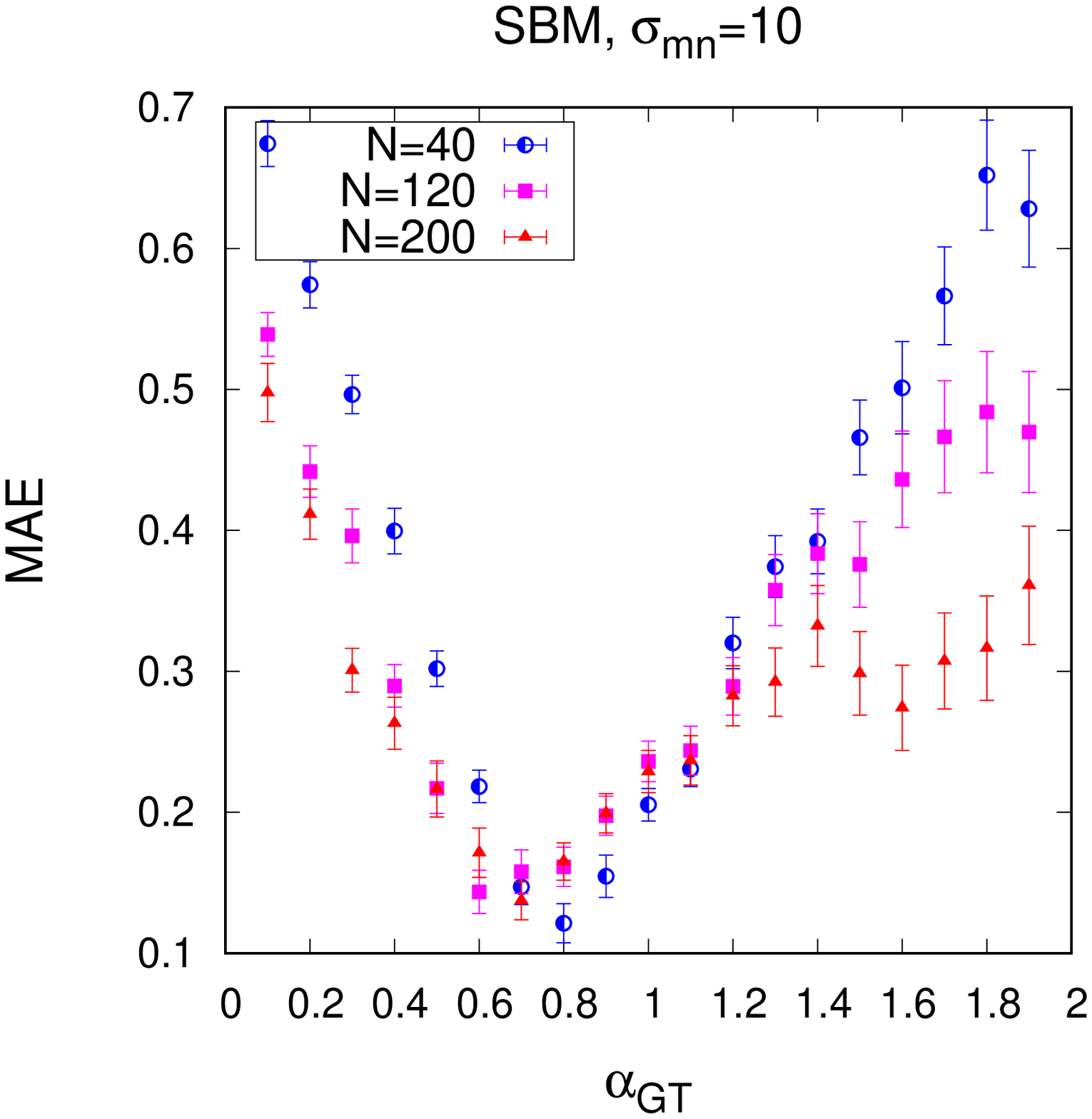} 
    \caption{MAE vs. $\alpha_{\rm GT}$ plot with the estimated MAE obtained from the analysis of $\tilde{N}=100$ 2-dimensional SBM trajectories for each $\alpha_{\rm GT}$. Each sub-plot shows the results from the analysis of SBM trajectories generated with different noise-strengths $\sigma_{\rm mn}$. The error-bars are the standard error on the mean.}
    \label{fig-mae-sbm-2d}
\end{figure}

\begin{figure}
    \includegraphics[angle=270,scale=0.35]{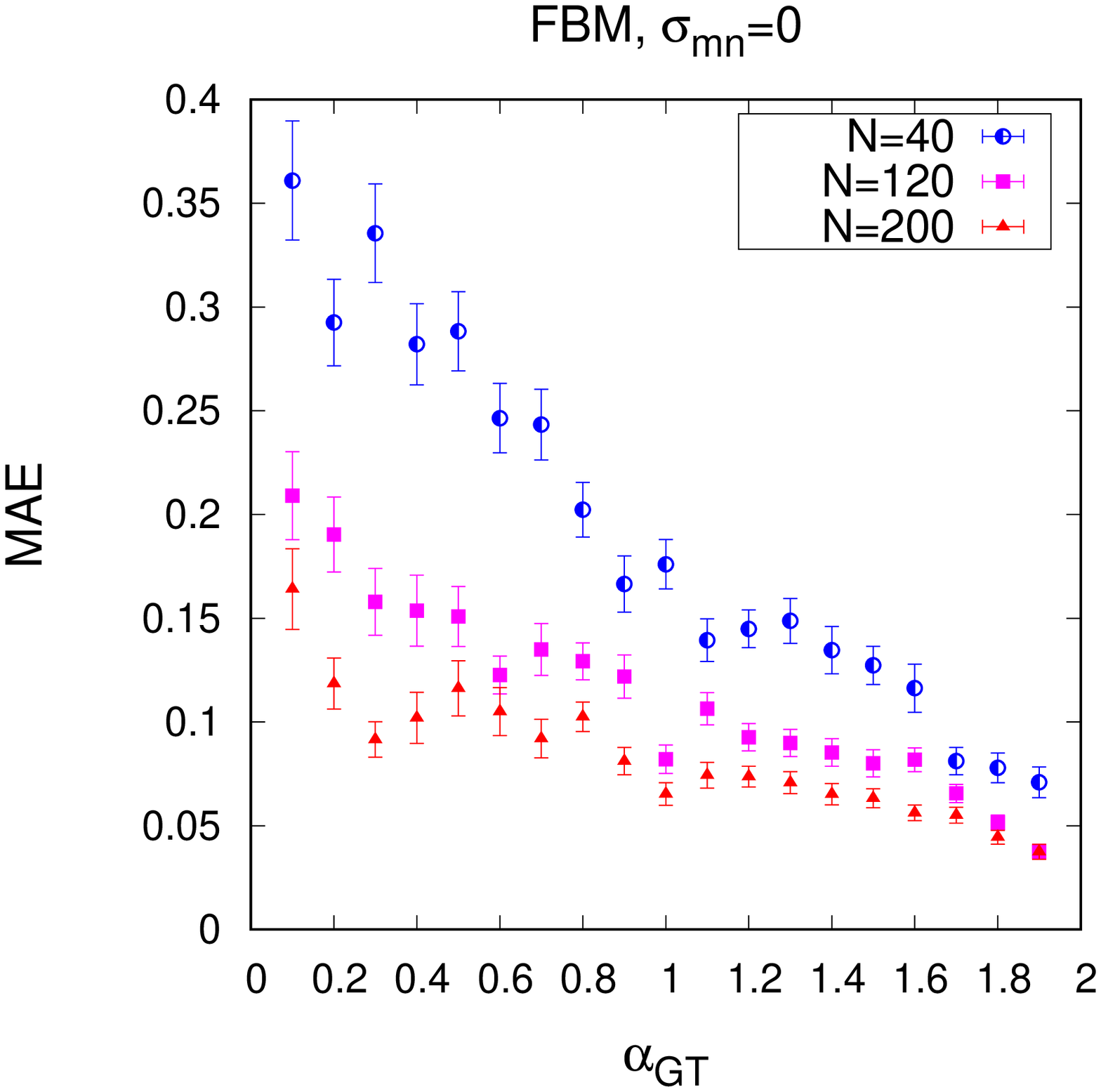} 
    \includegraphics[angle=270,scale=0.35]{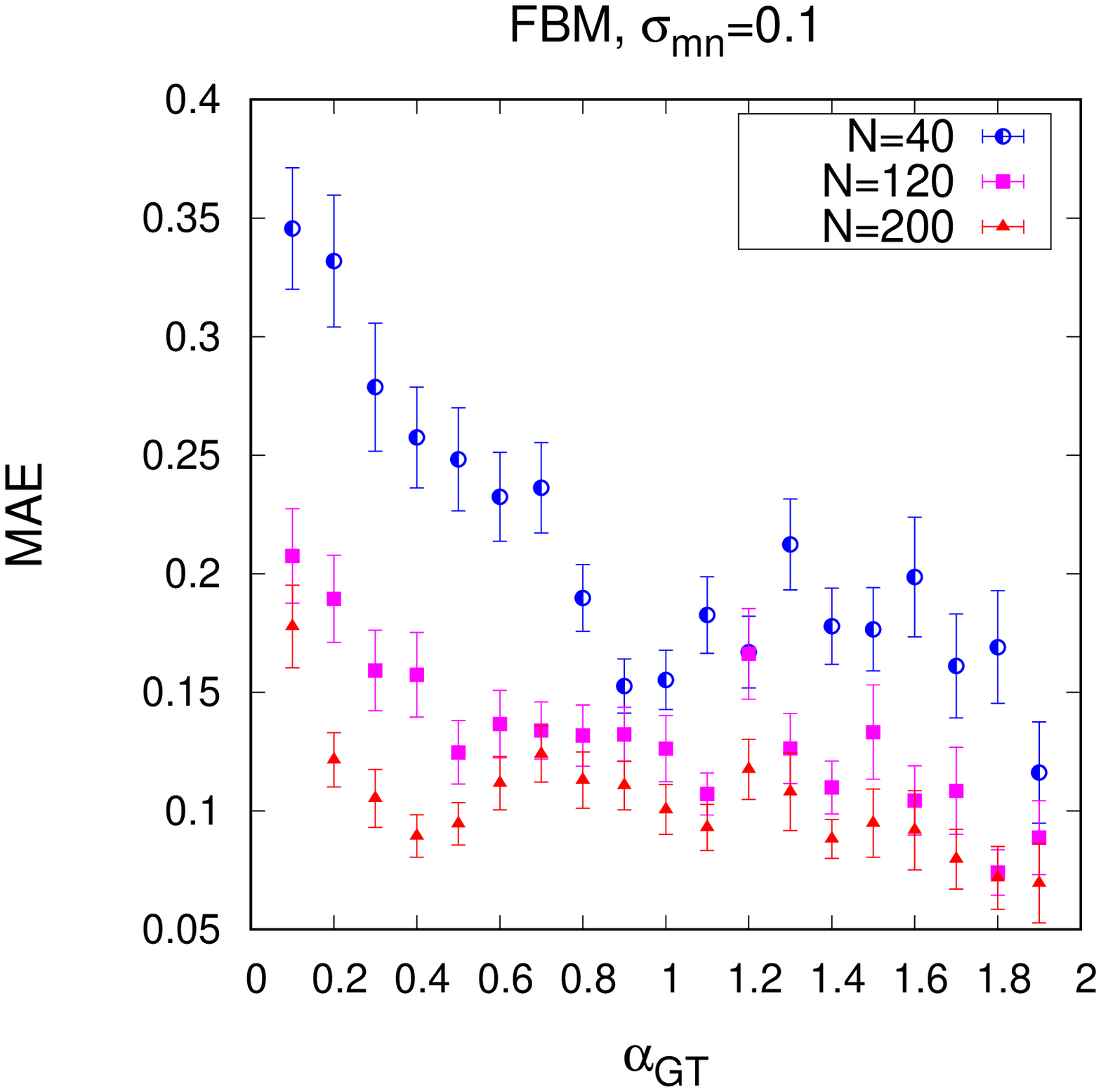}\\ 
    \includegraphics[angle=270,scale=0.35]{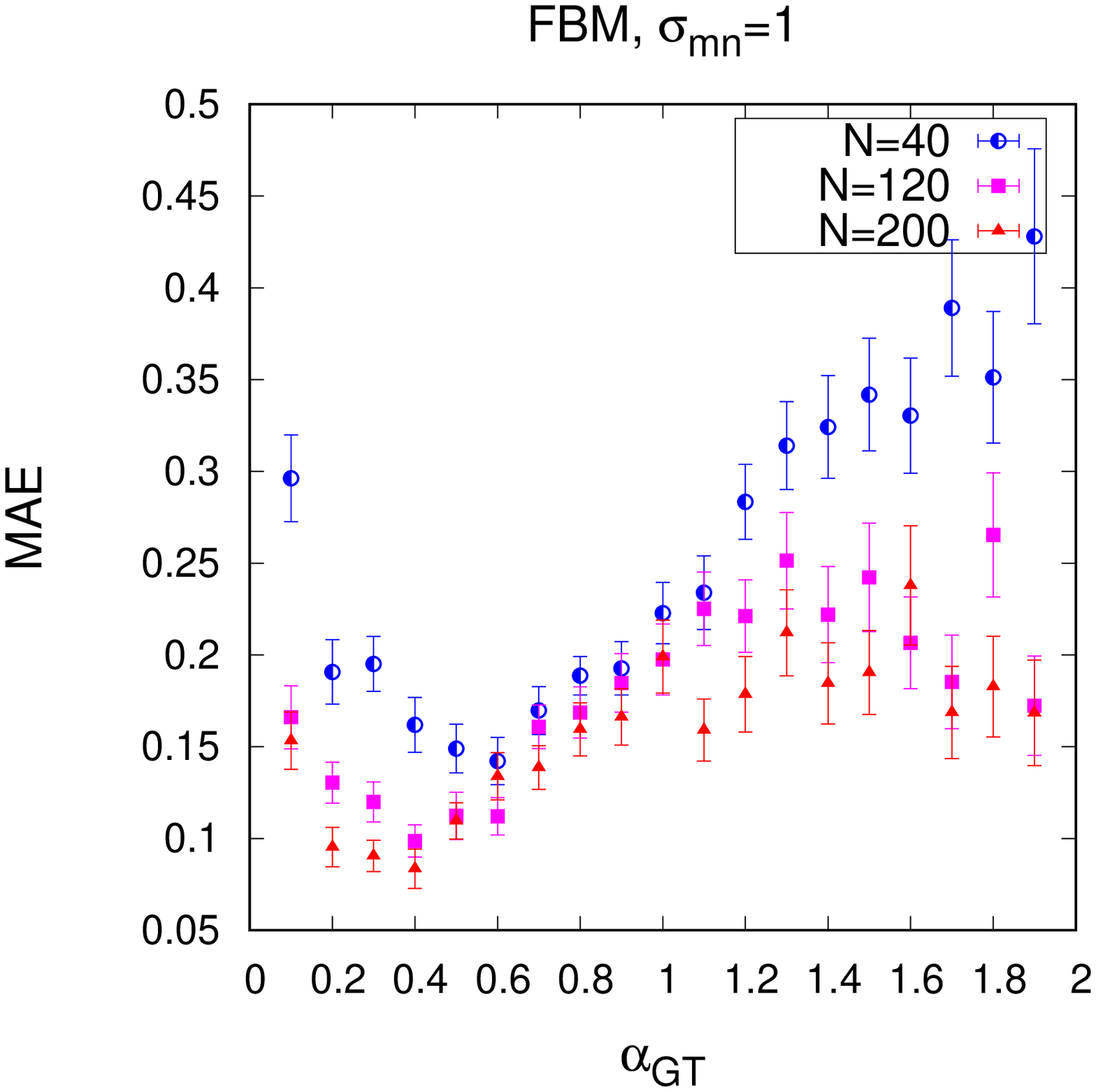} 
    \includegraphics[angle=270,scale=0.35]{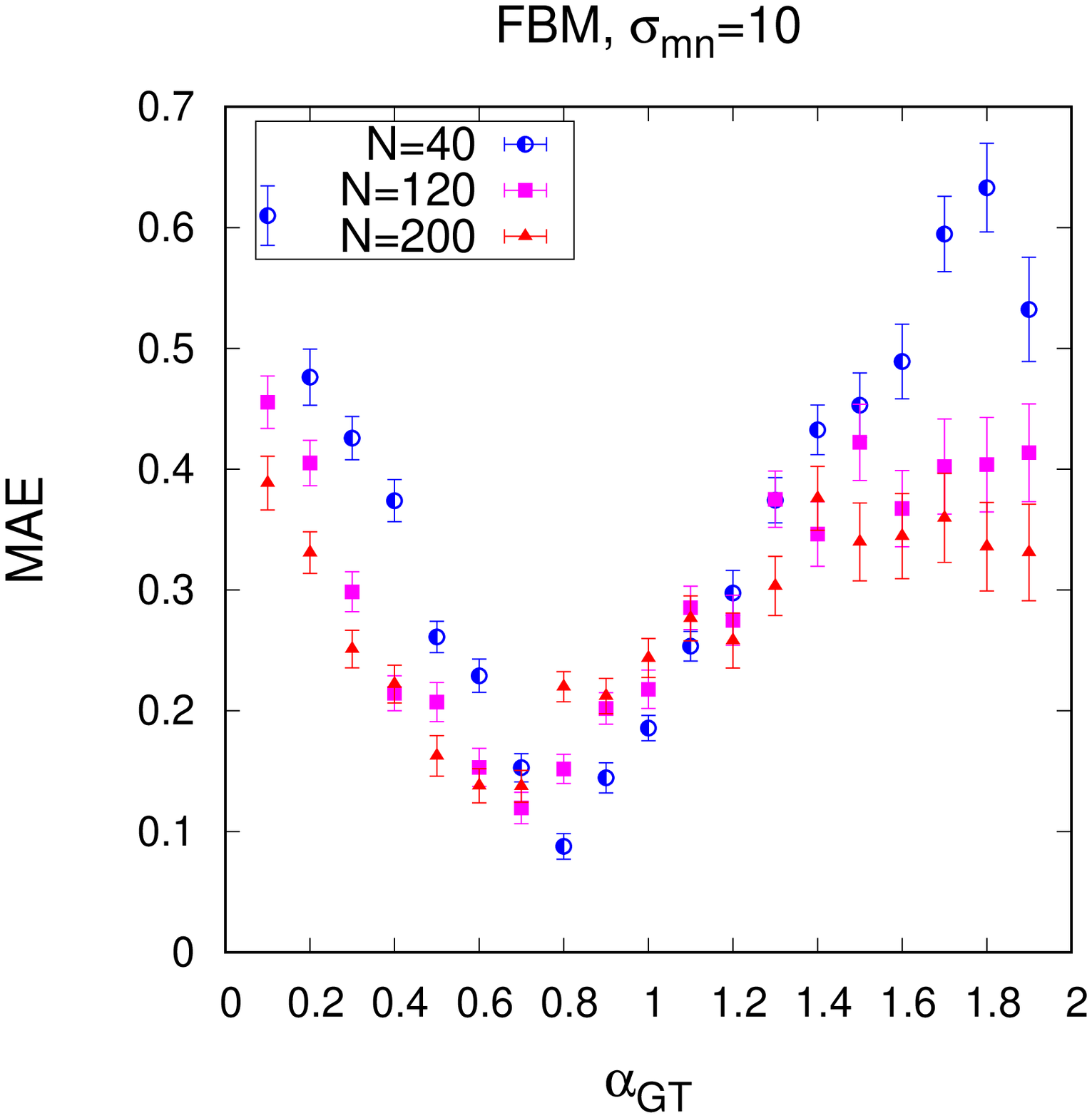} 
    \caption{MAE vs. $\alpha_{\rm GT}$ plot  with the estimated MAE obtained from the analysis of $\tilde{N}=100$ 2-dimensional FBM trajectories for each $\alpha_{\rm GT}$. Each sub-plot shows the results from the analysis of FBM trajectories generated with different noise-strengths $\sigma_{\rm mn}$. The error-bars are the standard error on the mean. }
    \label{fig-mae-fbm-2d}
\end{figure}

Figure \ref{fig-mae-sbm-2d} shows the results of MAE on the estimations of $\alpha$
as a function of $\alpha_{\rm GT}$ used to generate the 2-dimensional SBM trajectories. As expected, we see the estimations get better with the length $N$ of the trajectories. The analysis of noise-free trajectories ($\sigma_{\rm mn}=0$) shows that the estimation is better for high and low values of $\alpha_{\rm GT}$, whereas it gets relatively worse close to $\alpha_{\rm GT}=1$. A comparison of these results with those from the analysis of noisy trajectories shows interesting differences. While in general the analysis of noisy trajectories gives worse estimates of $\alpha$ as compared to the estimates from noise-free trajectories, the noise effects the estimates of very low and very high values of $\alpha$ much more than the values close to $\alpha_{\rm GT}=1$. We attribute part of the explanation of this to the enlarged possibility that a predicted $\alpha$ is far away from the ground truth when the ground truth is extreme. For instance, if the measurement noise strength is completely obscuring the actual position, we would expect the posterior on $\alpha$ to become equal to the uniform prior. Averaging over this uniform distribution with $0<\alpha<2$ we obtain $\langle {\rm MAE}\rangle = (\alpha_{\rm GT}-1)^2/2+1/2$, i.e., larger values in the cases where $\alpha_{\rm GT}$ is extreme. 
Figure \ref{fig-mae-sbm-1d} shows similar results from the analysis of 1-dimensional SBM trajectories.

On looking at the corresponding results from the analysis of 2-dimensional FBM trajectories in figure \ref{fig-mae-fbm-2d}, we find that for the cases of noise-free ($\sigma_{\rm mn}=0$) and relatively less noisy  ($\sigma_{\rm mn}=0.1$) trajectories, the MAE as a function of the $\alpha_{\rm GT}$
shows an interesting asymmetry. It decreases monotonically with increasing $\alpha_{\rm GT}$.  This can be understood from realising that the measurement noise is anti-persistent in nature, i.e., it induces anti-correlation between steps (see Eq. (\ref{eq:meanx})), and because subdiffusive FBM itself is anti-persistent, it seems reasonable that it becomes increasingly difficult to estimate
the anomalous diffusion exponent with increasing anti-persistence, i.e., for lower values of  $\alpha_{\rm GT}$. On increasing the noise-strength (the cases of $\sigma_{\rm mn}=1$ and $\sigma_{\rm mn}=10$), we recover the systematic increase in the values of MAE farther away from $\alpha_{\rm GT}=1$ in both directions like for SBM.
Figure \ref{fig-mae-fbm-1d} shows similar results from the analysis of 1-dimensional FBM trajectories.

\begin{figure}
    \includegraphics[angle=270,scale=0.35]{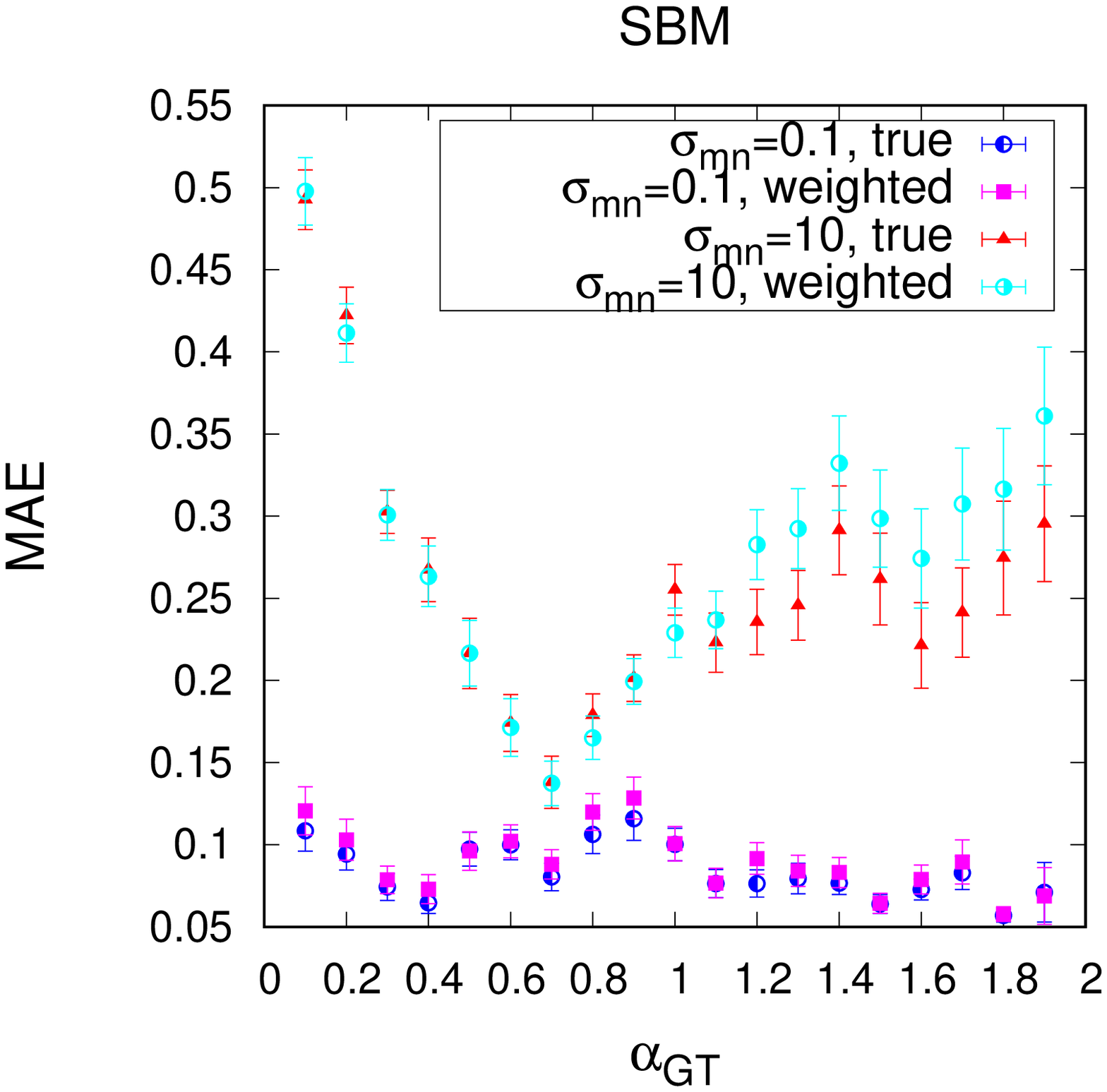}
     \includegraphics[angle=270,scale=0.35]{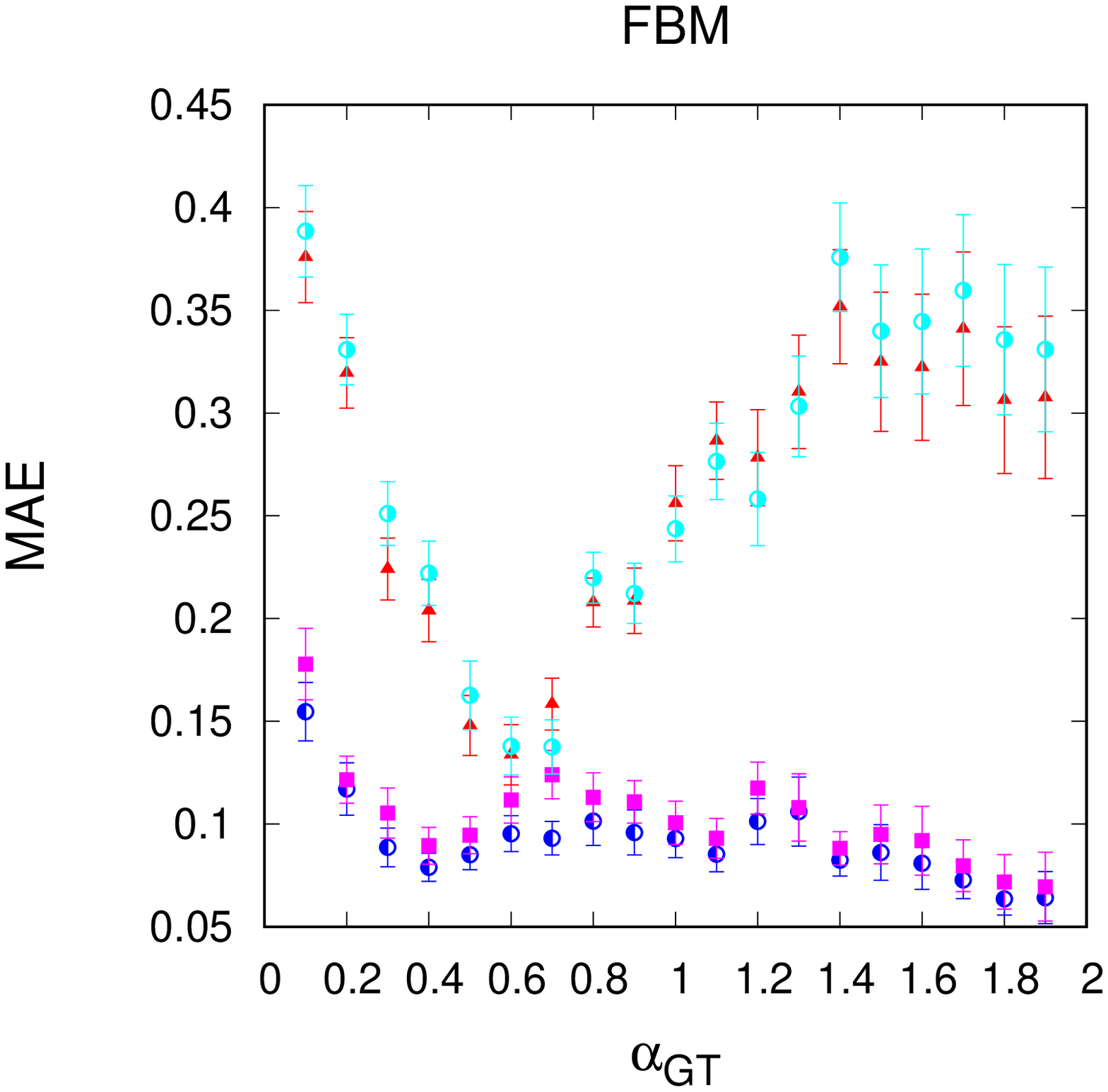}
    \caption{ Comparison of the MAE vs. $\alpha_{GT}$ plot between the estimates
    from the weighted average (labelled "weighted") and the estimates from the true model without
    considering model inference (labelled "true"). The error-bars are the standard error on the mean. 
    The labels are the same for the figure on the right as shown in the figure on the left.}
    \label{fig-mae-comp-2d}
\end{figure}

Taking cognisance of the fact that parameter estimation can be performed with or without conditioning on particular models, in figure \ref{fig-mae-comp-2d} we compare our estimation of $\alpha$ using Eq. (\ref{eq-alpha-inference})---where we considered the posterior median weighted with the model probabilities---with the inference using the posterior median of the true model. The figure on the left (right) shows results from the analysis of $\tilde{N}=100$ SBM (FBM) trajectories of length $N=200$ at each $\alpha_{GT}$. For both SBM and FBM trajectories, and for both $\sigma_{mn}=0.1$ and $\sigma_{mn}=10$, we find that using the posterior of the true model does not improve the MAE significantly. 

\begin{figure}
    \includegraphics[angle=270,scale=0.35]{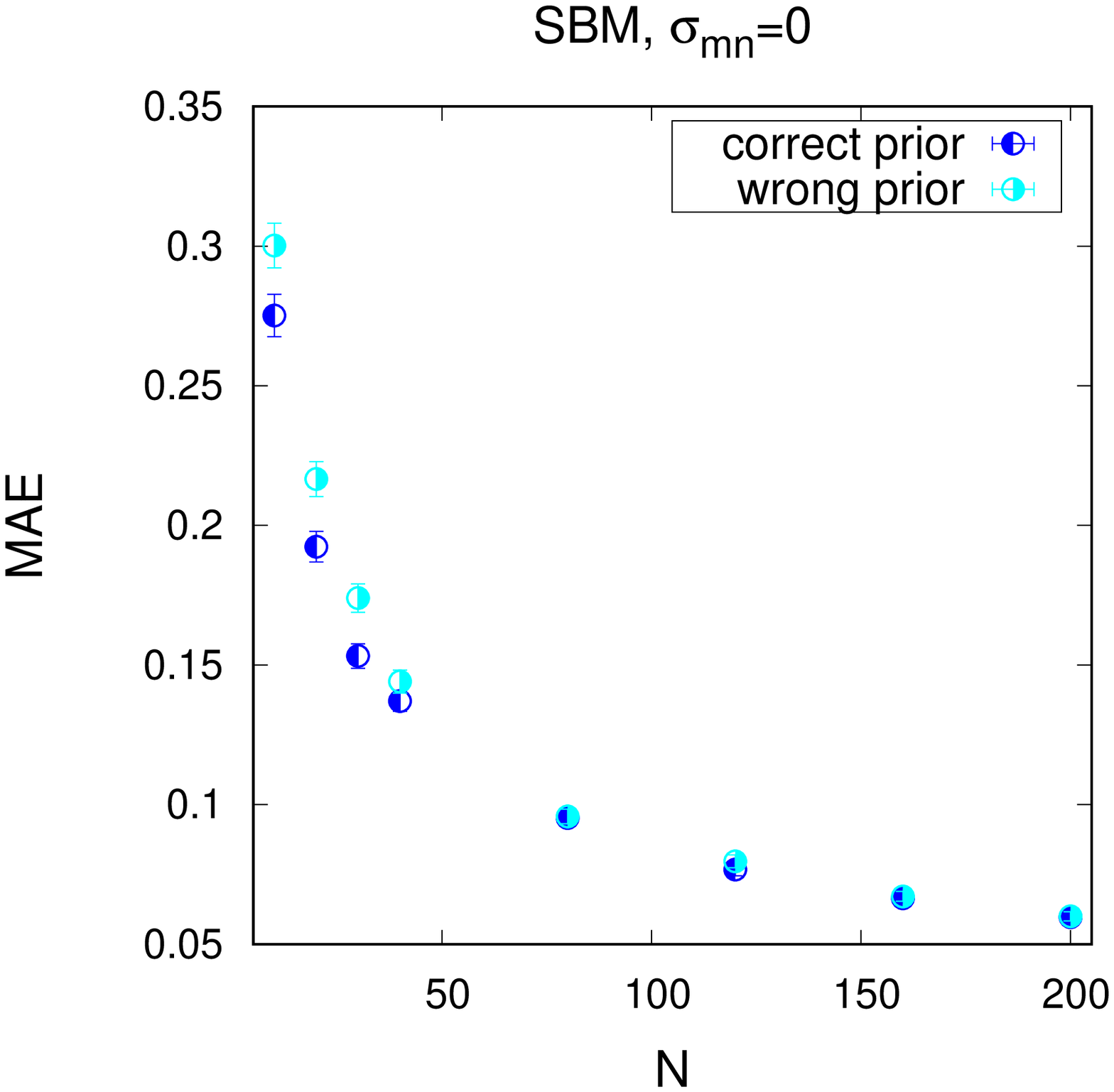}
    \includegraphics[angle=270,scale=0.35]{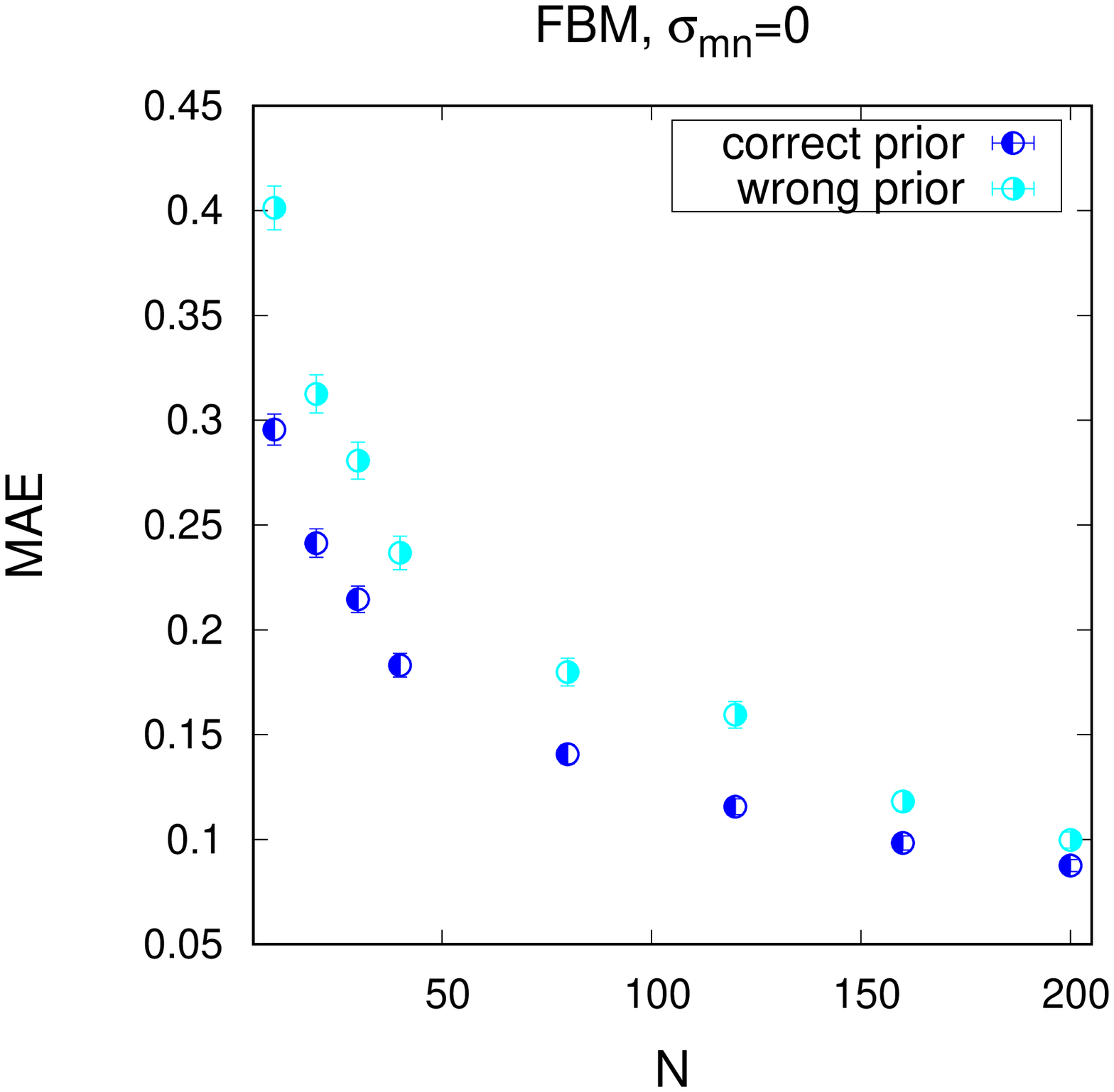}
    \caption{MAE vs. $N$ plot  with the estimated MAE obtained from the analysis of $\tilde{N}=1000$ 2-dimensional SBM (left) and  FBM (right) trajectories. The results labelled as "wrong prior" were inferred using a linear prior on $\alpha$. The error-bars are the standard error on the mean. }
    \label{fig-mae-wp-2d}
\end{figure}

In figure \ref{fig-mae-wp-2d} we show what effect choosing a wrong prior on $\alpha$ has on its
estimation. The results labelled "wrong prior" correspond to a linear prior on $\alpha$ whereas
those labelled "correct prior" correspond to a uniform prior (see section \ref{sec-priors}). The figure on the left shows the MAE vs. $N$ plot from the analysis of $\tilde{N}=1000$ noise-free SBM trajectories for each $N$. Each trajectory was generated with a $\alpha_{\rm GT}$ drawn from a uniform distribution in the range $0<\alpha_{\rm GT}<2$. The figure on the right shows the corresponding results from the analysis of $\tilde{N}=1000$ noise-free FBM trajectories for each $N$. In both cases we observe that choosing the correct prior results in a better estimate of $\alpha$ at low $N$. The estimations get better with longer trajectories and converges for the two choices of prior. Since the increments of FBM are correlated there are effectively fewer independent data points and it makes sense that the convergence for the two choices of prior happens more slowly with increasing $N$ in that case.

\subsection{Quantification of model selection}
\label{subsec:model}

\begin{figure}
    \includegraphics[angle=270,scale=0.35]{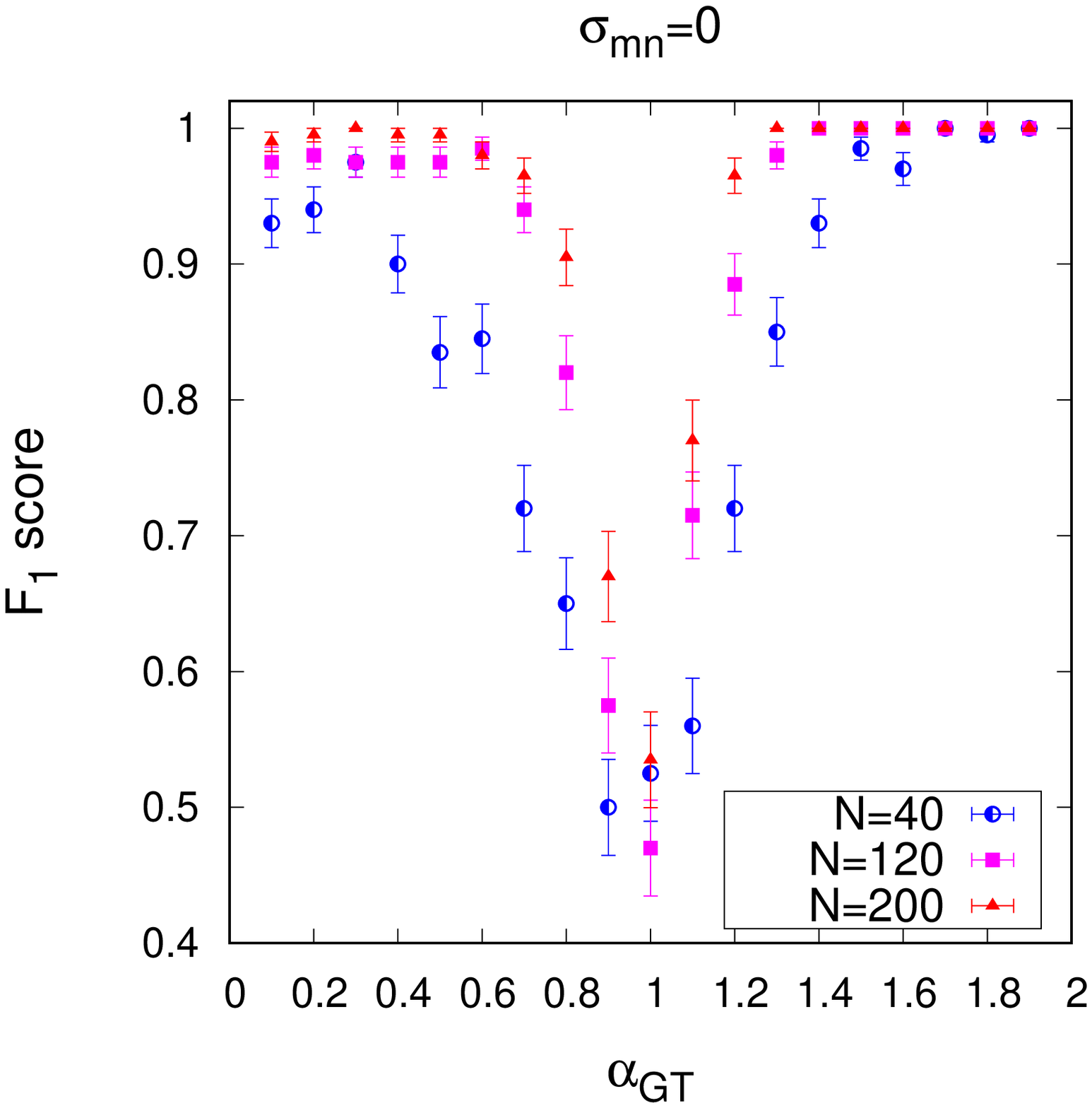}
    \includegraphics[angle=270,scale=0.35]{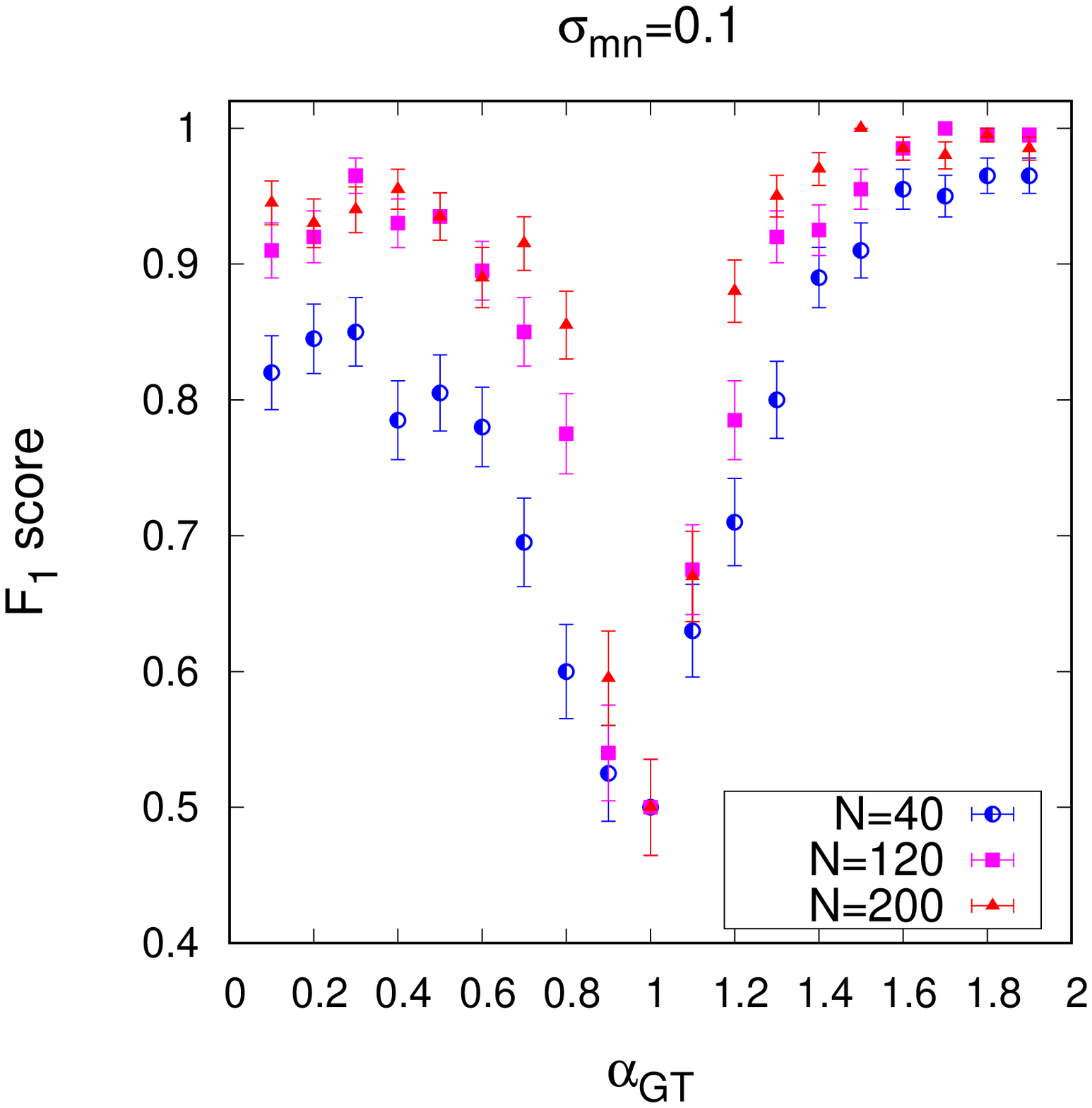}\\ 
    \includegraphics[angle=270,scale=0.35]{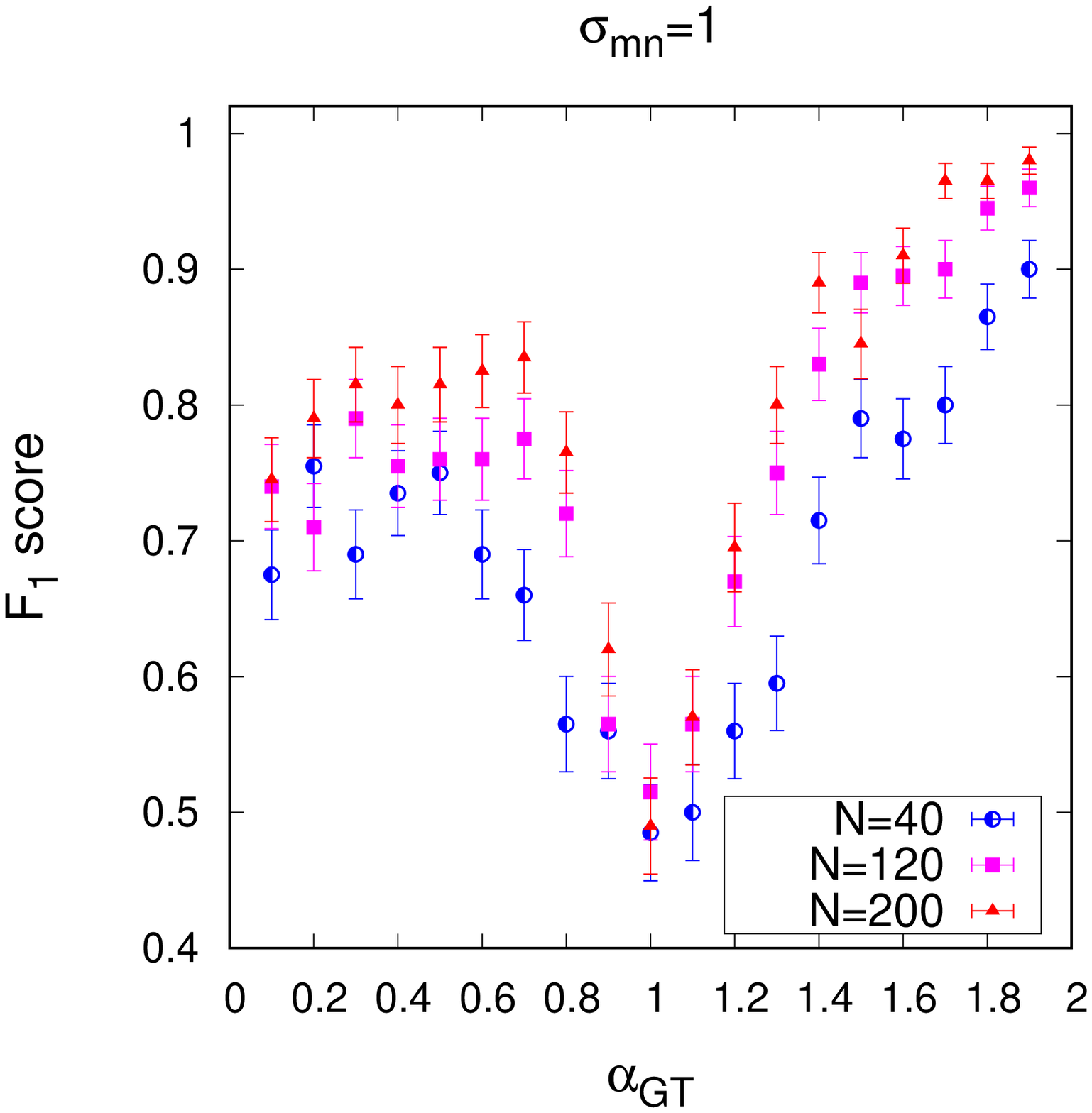} 
    \includegraphics[angle=270,scale=0.35]{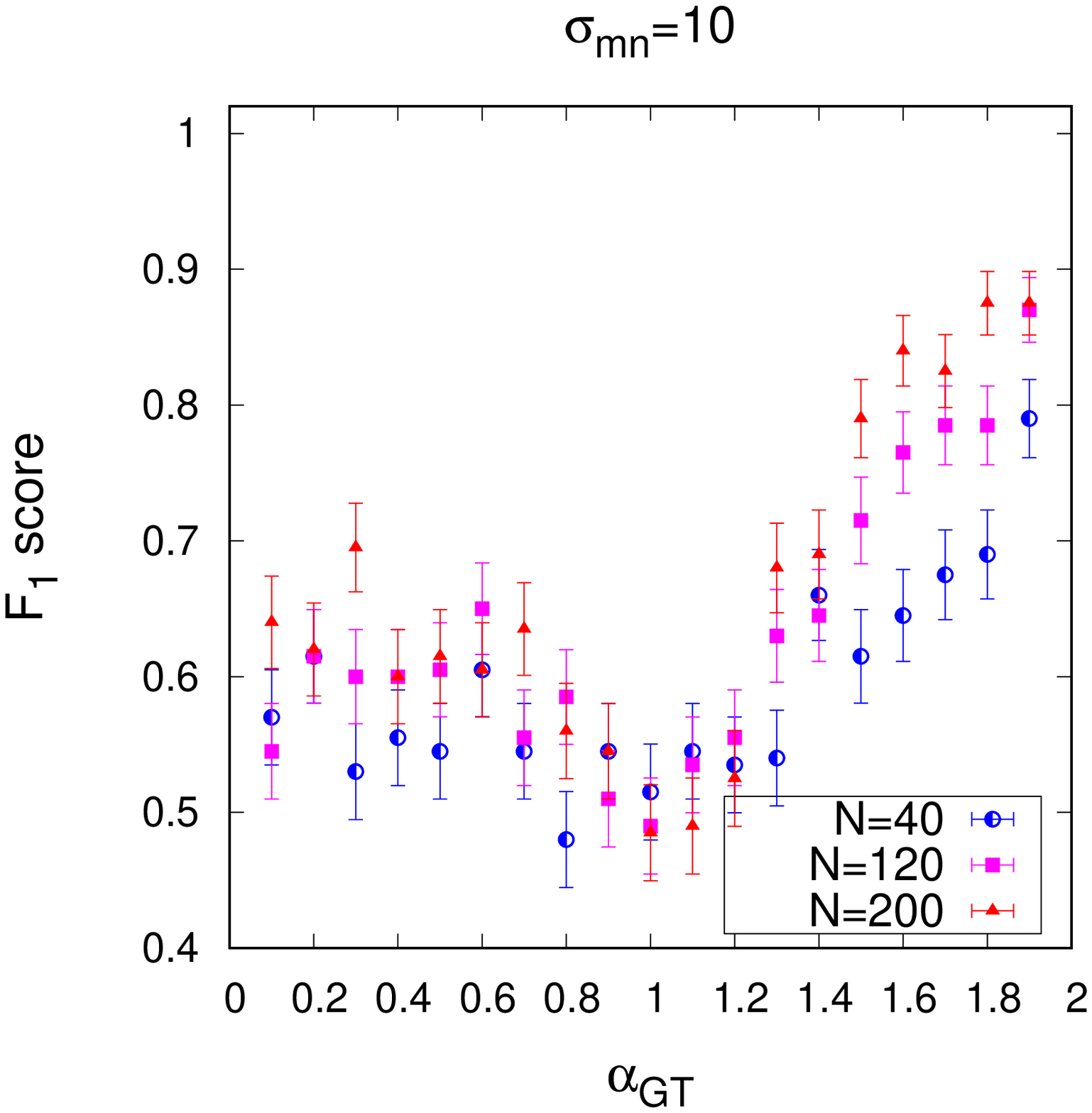}
    \caption{$F_1$ score vs. $\alpha_{\rm GT}$ plot from the analysis of 2-dimensional FBM  and SBM trajectories. 
    The $F_1$ scores at each $\alpha_{\rm GT}$ are from the analysis of $\tilde{N}=100$ SBM and $\tilde{N}=100$ FBM trajectories generated with that particular $\alpha_{\rm GT}$.
    The error-bars are the standard error on the mean for a Bernoulli random variable.}
    \label{fig-f1-2d}
\end{figure}

To quantify the model comparison results we use the $F_1$ metric, which is defined as \cite{andi21}
\begin{equation}
\label{eq-f1}
    F_1=\frac{\textrm{True positive}}{{N_{\rm total}}},
\end{equation}
where \textit{true positive} denotes the total number of trajectories assigned to the correct model after model comparison and ${N_{\rm total}}$ denotes the total number of trajectories considered. \textchg{red}{We assign each trajectory $\Delta\vek{x}_N$ to the model $M_i$ that has the highest value of the posterior model probability $P(M_i|\Delta\vek{x}_N)$ as defined in Eq. (\ref{eq-model-prob}).}
Except when we compare the effect of different choices of priors, we consider two models (FBM and SBM) with $\tilde{N}=100$ trajectories each, and therefore with ${N_{\rm total}}=200$. Just as in the case of parameter estimation, to study the effect of different choices of priors on $\alpha$ we consider the $F_1$ score from FBM and SBM trajectories but with $\tilde{N}=1000$ trajectories each, and therefore with ${N_{\rm total}}=2000$. Note that $F_1$ is always in the range $0 \leq F_1 \leq 1$ with a high value of $F_1$ corresponding to better model prediction. The error on the estimate of $F_1$ is obtained as the standard error on the mean of a Bernoulli random variable. This is estimated as $\sqrt{p(1-p)/N_{\rm total}}$ where $p$ is the fraction of trajectories for which the model prediction is correct.   

\begin{figure}
    \includegraphics{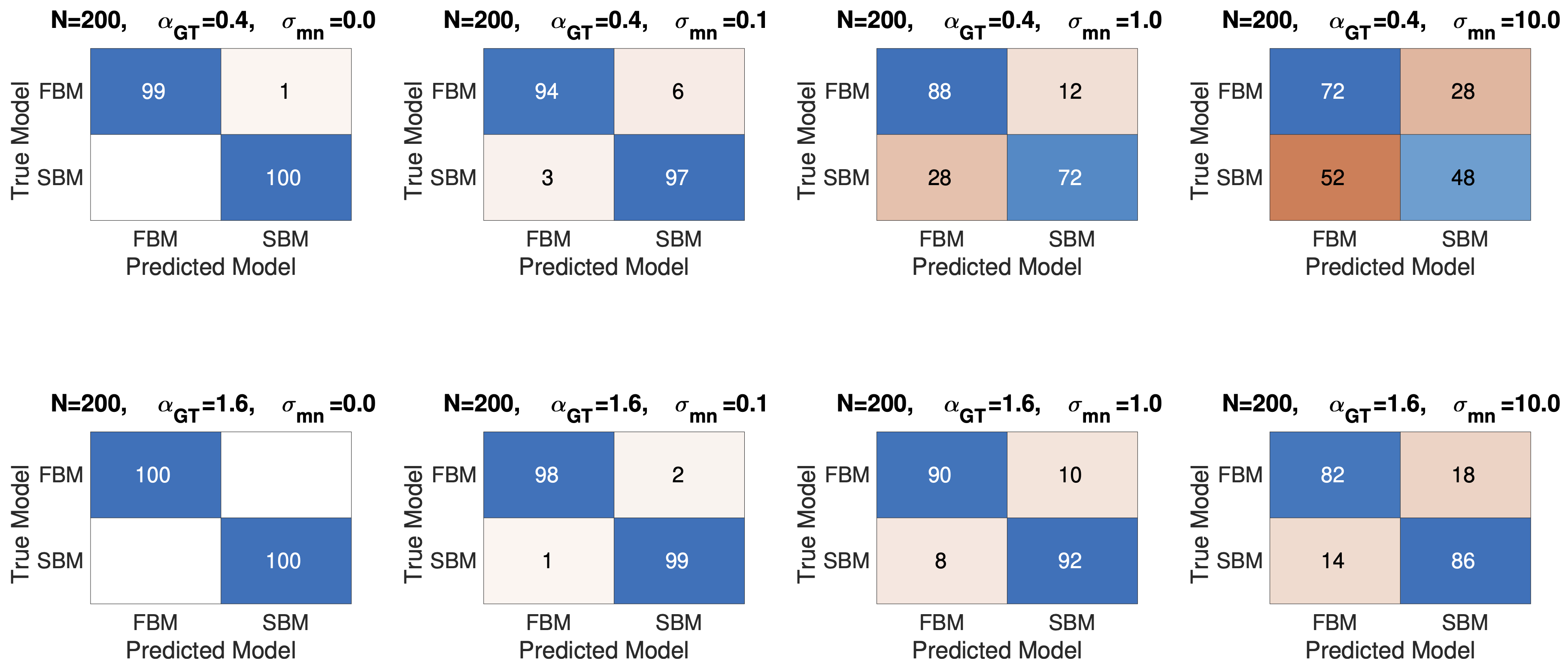}
    \caption{Plot of the confusion matrix highlighting the asymmetry between subdiffusive and superdiffusive cases during model inference. For the subdiffusive case ($\alpha_{\rm GT}=0.4$) shown in the upper panel, with increasing noise-strength model inference gets asymmetrically worse, with more SBM trajectories incorrectly inferred as FBM than FBM trajectories inferred as SBM. The lower panel shows that for the superdiffusive case ($\alpha_{\rm GT}=1.6$), model inference gets more symmetrically worse with increasing noise-strength.  }
    \label{fig-confusion-2d}
\end{figure}

Figure \ref{fig-f1-2d} shows the $F_1$ vs. $\alpha_{\rm GT}$ plot for different lengths of the trajectories, $N$ and each sub-plot shows results from the analysis of trajectories generated with a specific noise-strength $\sigma_{\rm mn}$. For all noise-strengths, at $\alpha_{\rm GT}=1$, $F_1 \approx 0.5$ which is expected because at this value of $\alpha_{\rm GT}$ both FBM and SBM are identical to Brownian motion and therefore should be indistinguishable.  In addition to the expected result that the $F_1$ score gets better with increasing length of the trajectories, and worse with increasing strength of the measurement noise,  we see some interesting features. The analysis of noise-free trajectories ($\sigma_{\rm mn}=0$) gives $F_1$ scores more or less symmetric around $\alpha_{\rm GT}=1$, with a score getting systematically better
with $\alpha_{\rm GT}$ farther away from $\alpha_{\rm GT}=1$ in both directions. The addition of measurement noise breaks this symmetry, with the $F_1$ score getting profoundly lower with increasing noise-strength for subdiffusive trajectories as compared to superdiffusive trajectories. This can be understood in terms of the measurement noise being anti-persistent in nature. With increasing noise-strength the SBM trajectories develop anti-persistent characteristics (see Eq. (\ref{eq:meanx})) making them difficult to distinguish from subdiffusive FBM trajectories which are inherently anti-persistent.  This should lead a higher number of SBM trajectories incorrectly predicted as being FBM trajectories in comparison to the number of FBM trajectories falsely detected as SBM trajectories. Moreover, this effect should occur in the subdiffusive case but not in the superdiffusive case where noise-free FBM is persistent, not anti-persistent. This is exactly what we see in figure \ref{fig-confusion-2d}, which shows the comparison of model inference via the confusion matrix plot. The upper panel shows the results from subdiffusive ($\alpha_{\rm GT}=0.4$) SBM and FBM trajectories while the lower panel shows the results from superdiffusive ($\alpha_{\rm GT}=1.6$) SBM and FBM trajectories. The trajectories were generated with $N=200$ and fixed $\sigma_{\rm mn}$ (as specified on the title of each sub-plot) for both cases. The model prediction is very good for both cases, and for both FBM and SBM trajectories in the case of small noise-strengths ($\sigma_{\rm mn}=0$ and $\sigma_{\rm mn}=0.1$).  However, for  large noise-strengths ($\sigma_{\rm mn}=1$
and $\sigma_{\rm mn}=10$) we clearly see an asymmetry in the model prediction between subdiffusive FBM and SBM trajectories. With increasing noise-strength, indeed a higher number of SBM trajectories are predicted as FBM whereas the model prediction for FBM trajectories remain reasonably well. This asymmetry in the model prediction results is not seen in the superdiffusive case where, with increasing noise-strength, the incorrect attribution of FBM trajectories to SBM and vice-versa increases almost symmetrically, as expected.

\begin{figure}
    \centering
    \includegraphics[angle=270,scale=0.35]{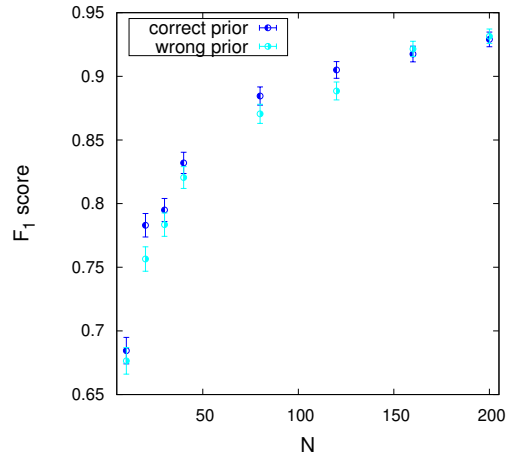}
    \caption{$F_1$ score vs. $N$ plot from the analysis of 2-dimensional noise-free SBM and FBM trajectories.
    The $F_1$ score at each $N$ are from the analysis of $\tilde{N}=1000$ SBM and $\tilde{N}=1000$ FBM trajectories generated with that particular $N$ and a $\alpha_{\rm GT}$ chosen from a uniform distribution in the range $0<\alpha<2$. The results labelled as "wrong prior" were inferred using a linear prior on $\alpha$. 
    The error-bars are the standard error on the mean for a Bernoulli random variable.   }
    \label{fig-f1-wp-2d}
\end{figure}

Figure \ref{fig-f1-wp-2d} shows what effect using a wrong prior on $\alpha$ has on model prediction. We consider the same data sets used in figure \ref{fig-mae-wp-2d}  where we discussed the effect of a wrong prior on parameter estimation. We observe a systematic decrease in $F_1$ score on using a wrong prior on $\alpha$ for low $N$.
However the score itself is very close to the one obtained using the correct prior, and the differences between the two choices of prior vanishes with more data. 

\subsection{CTRW trajectories}
\begin{figure}
    \includegraphics[angle=270,scale=0.35]{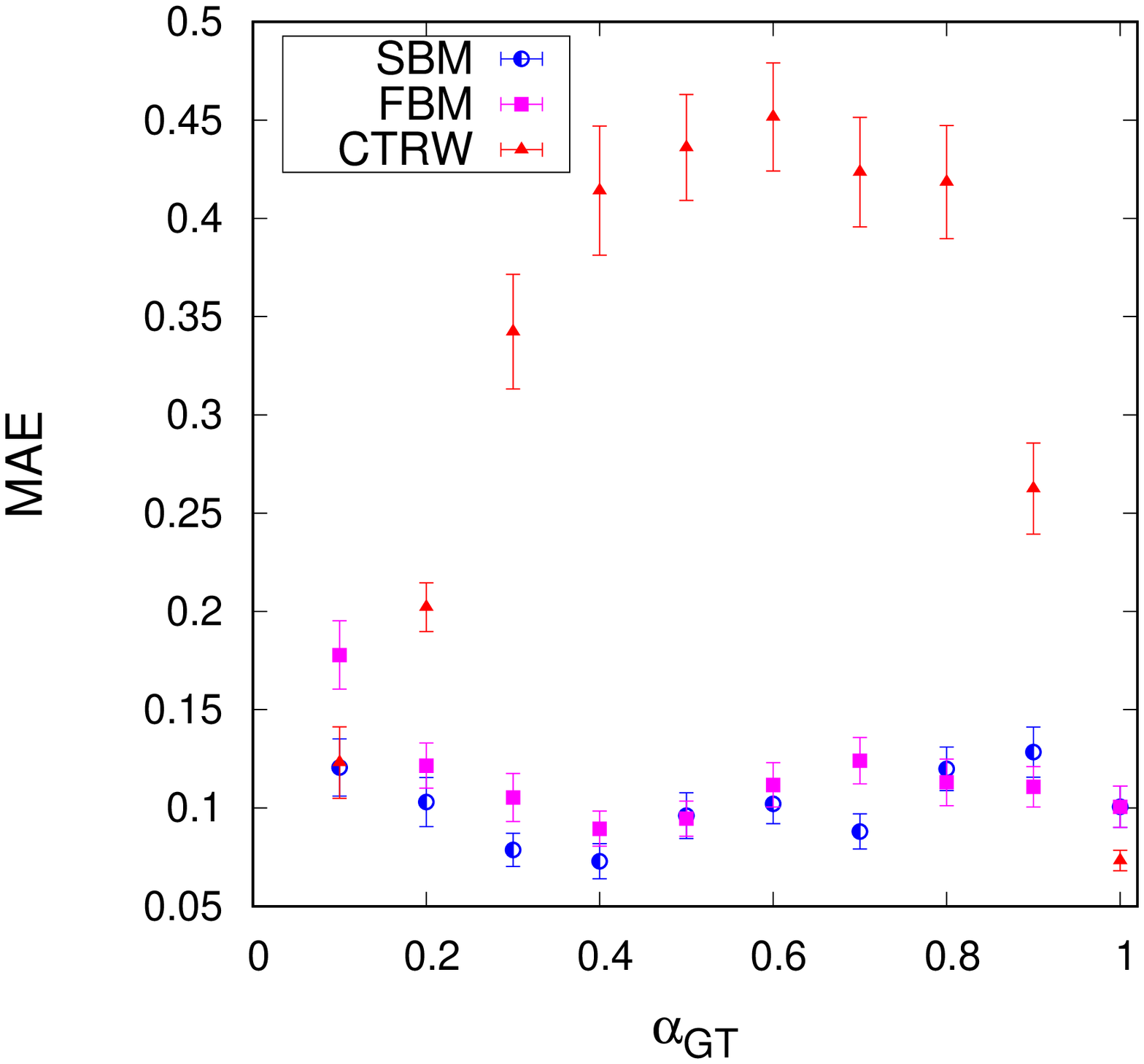}
        \includegraphics[angle=270,scale=0.33]{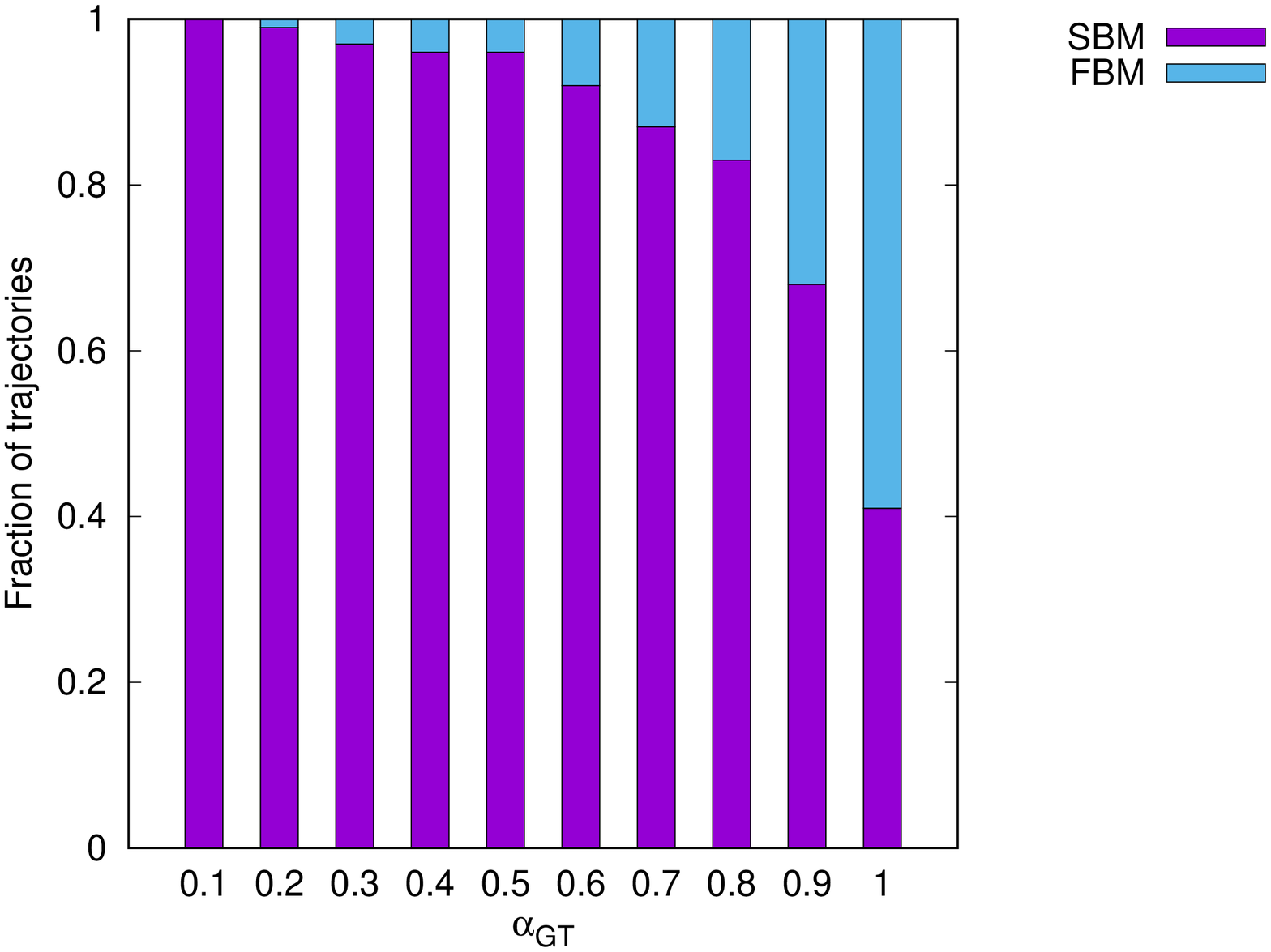}
    \caption{Left: MAE vs. $\alpha_{GT}$ plot  from 2-dimensional CTRW trajectories of $N=200$ points with added measurement noise of strength $\sigma_{mn}=0.1$ (labelled "CTRW"). $\tilde{N}=100$ trajectories were generated for each $\alpha_{GT}$. The estimates from FBM and SBM trajectories (labelled "FBM" and "SBM" respectively)---with $\sigma_{mn}=0.1$ and $N=200$ points from figures \ref{fig-mae-sbm-2d} and \ref{fig-mae-fbm-2d}---are presented for comparison. All the MAE estimates are done using Eq. (\ref{eq-mae}) and Eq. (\ref{eq-alpha-inference}) with two models, FBM and SBM. Right: Model inference using FBM and SBM models on 2-dimensional CTRW trajectories generated with added measurement noise of strength $\sigma_{mn}=0.1$. $\tilde{N}=100$ trajectories were generated for each $\alpha_{GT}$.  }
    \label{fig-ctrw-2d}
\end{figure}
The results from Bayesian inference obviously depend on the list of models considered in the analysis. We highlight this in figure \ref{fig-ctrw-2d} where we present results from the analysis of simulated 2-dimensional continuous time random walk (CTRW) trajectories of length $N=200$, with an added measurement noise of strength $\sigma_{mn}=0.1$. CTRW has been used to model numerous experiments exhibiting anomalous diffusion\cite{krapf11,weitz04,chaikin11,swinney93}. The chosen CTRW model is a renewal process with Gaussian jump-lengths with an asymptotic power law waiting time between the jumps\cite{scher1975,montroll1969}.
The exact probability density function of the waiting time is a one-sided $\alpha$-stable distribution \cite{scher1975,montroll1969}, where $0 \leq \alpha \leq 1$ is also the anomalous diffusion exponent \cite{scher1975,montroll1969}.
The CTRW trajectories were simulated following ref.\cite{hans07}.\footnote{We choose $\Delta s=0.01$ and $\Delta t =1$ in the simulations. See ref.\cite{hans07} for details.}
The Bayesian inference was done with SBM and FBM models.
The MAE estimates using Eq. (\ref{eq-mae}) and Eq. (\ref{eq-alpha-inference}) clearly show the bad performance  for the CTRW trajectories---except when $\alpha_{\rm GT}$ is very small or close to 1---as compared with the estimate from SBM and FBM trajectories with the same noise strength 
of $\sigma_{mn}=0.1$ and number of points $N=200$. The results from model inference with FBM and SBM models on CTRW trajectories are also interesting, albeit expected.  The  right panel in figure \ref{fig-ctrw-2d} shows the model inference results on the same CTRW trajectories which were analysed to generate the MAE vs. $\alpha_{\rm GT}$ plot on the left. It shows the fraction of trajectories inferred as most probably SBM (FBM) as the proportion of purple (blue) colour in a bar at a given $\alpha_{\rm GT}$. We clearly see that most of the CTRW trajectories are inferred as SBM unless $\alpha_{\rm GT} \to 1$ when CTRW, SBM and FBM all converge towards BM and thus towards becoming indistinguishable. In this case, we see a comparable fraction of trajectories inferred as FBM. Indeed, the inference of CTRW trajectories as SBM rather than FBM---in the absence of CTRW in the list of models considered---is expected. This is because SBM  can be considered
as a homogenised (mean-field) approximation to CTRW and therefore is its close relative, as was argued in \cite{thiel2014}. \textchg{red}{In particular, the increments of FBM are stationary and anti-persistent, while this is not the case for both CTRW and SBM.}

\section{Conclusion}\label{sec:concl}

We implemented Bayesian inference for SBM, and tested the procedure in combination with FBM on synthetic data. The results obtained in these tests are an upper limit on how well any inference method can perform, since the procedure is singled out as optimal by Bayes' theorem and decision theory. In this connection we note that the issue of falsely inferring many SBM trajectories as FBM (see our figure 7) also applies for the top machine learning method in the AnDi competition (see figure 3e in \cite{andi21}). For the computational efficiency of the Bayesian approach it is important that the hidden variables (e.g., measurement noise) can be integrated out analytically, which can be achieved for SBM and FBM. For models where this analytic integration is too difficult, one can instead attempt to integrate the hidden variables numerically or with Monte Carlo simulations. However, if this is too expensive computationally, one might resort to machine learning methods as an alternative for both parameter and model inference \cite{andi21}. \textchg{red}{But note that while a Bayesian analysis automatically provides uncertainty estimates for the inferred parameters, this is not the case for neural networks \cite{khosravi11}. This is because the result of a Bayesian analysis is a posterior distribution for the parameters, unlike point estimates from a neural network.}

Bayesian inference is sometimes criticised for being subjective, because a prior has to be chosen, which specifies the beliefs in the models and parameters prior to any data being revealed. The problem diminishes, as more data is obtained, something we also tested here by comparisons with results from using a wrongly skewed prior on the $\alpha$-parameter. We would like to point out, that machine learning methods suffer from the same issue of having to choose a prior distribution, since distributions of models and parameters have to be chosen for the data that is used for training these methods.

For real experimental data any simple mathematical model will not be an exact match of the underlying process. Thus the methods discussed in this article cannot stand alone, but has to be supplemented by model checking \cite{gelman13} (also called goodness-of-fit tests) to check how well the inferred model describes the observed data, and possibly, if too big discrepancies are found, improved models have to be designed and the inference process reiterated with these models. 

\ack
S Thapa acknowledges support in the form of a Sackler postdoctoral fellowship 
and funding from the Pikovsky-Valazzi matching scholarship, Tel Aviv University. J-H Jeon acknowledges support from the National Research Foundation (NRF) of Korea (No.~2020R1A2C4002490). R Metzler acknowledges funding from the German Science Foundation (DFG, grant n. ME 1525/12-1) and the Foundation for Polish Science (Fundacja na rzecz Nauki Polskiej, FNR) within an Alexander von Humboldt Honorary Polish Research Scholarship.

\clearpage
\appendix
\section{Additional figures (from the analysis of 1-dimensional trajectories)}
\begin{figure}[!h]
    \includegraphics{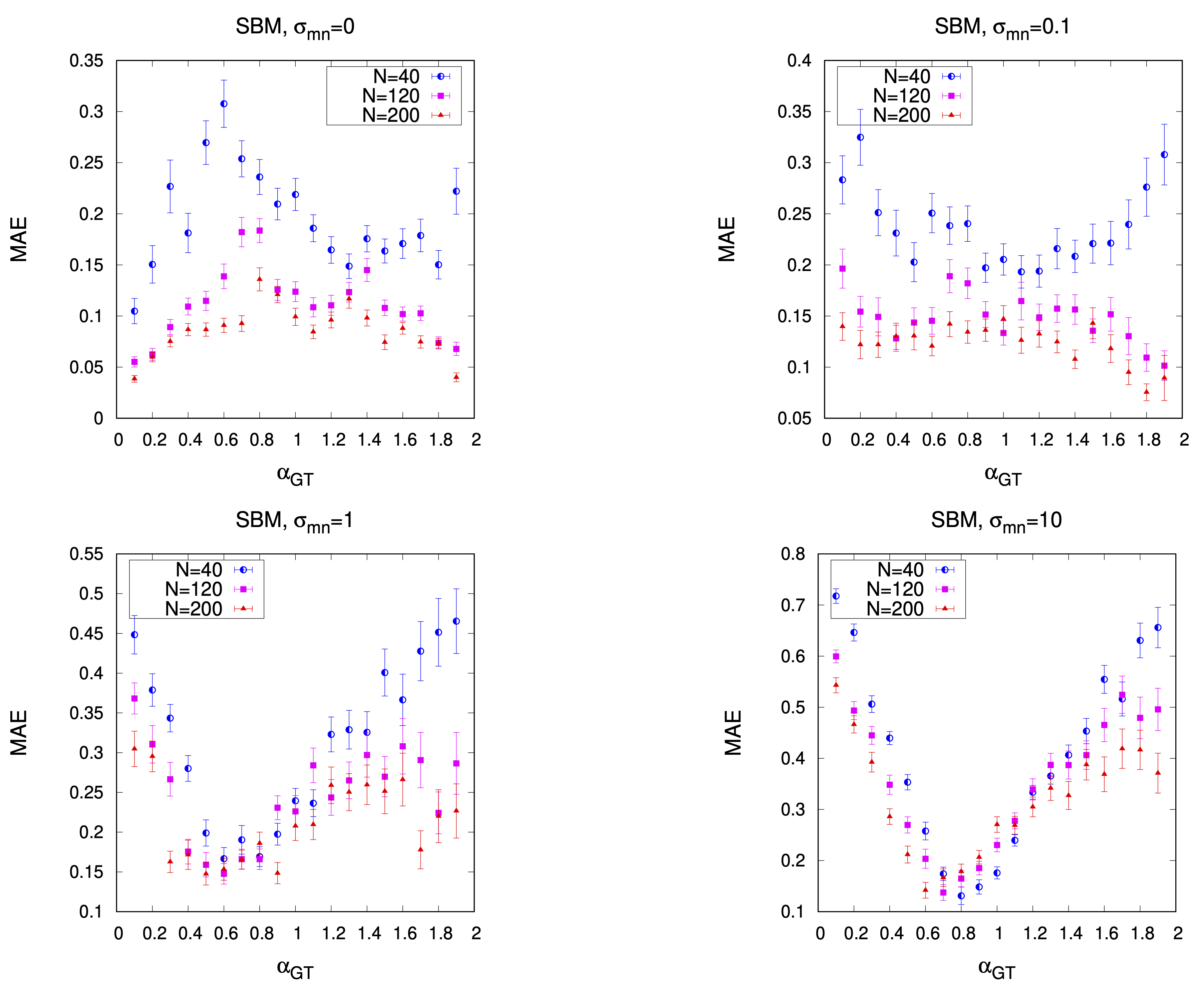} 
    \caption{MAE vs. $\alpha_{\rm GT}$ plot with the estimated MAE obtained from the analysis of $\tilde{N}=100$ 1-dimensional SBM trajectories for each $\alpha_{\rm GT}$. Each sub-plot shows the results from the analysis of SBM trajectories generated with different noise-strengths $\sigma_{\rm mn}$. The error-bars are the standard error on the mean. This figure is equivalent to figure \ref{fig-mae-sbm-2d}, except that it is for 1-dimensional trajectories instead of 2-dimensional ones.}
    \label{fig-mae-sbm-1d}
\end{figure}

\begin{figure}
    \includegraphics[angle=270,scale=0.35]{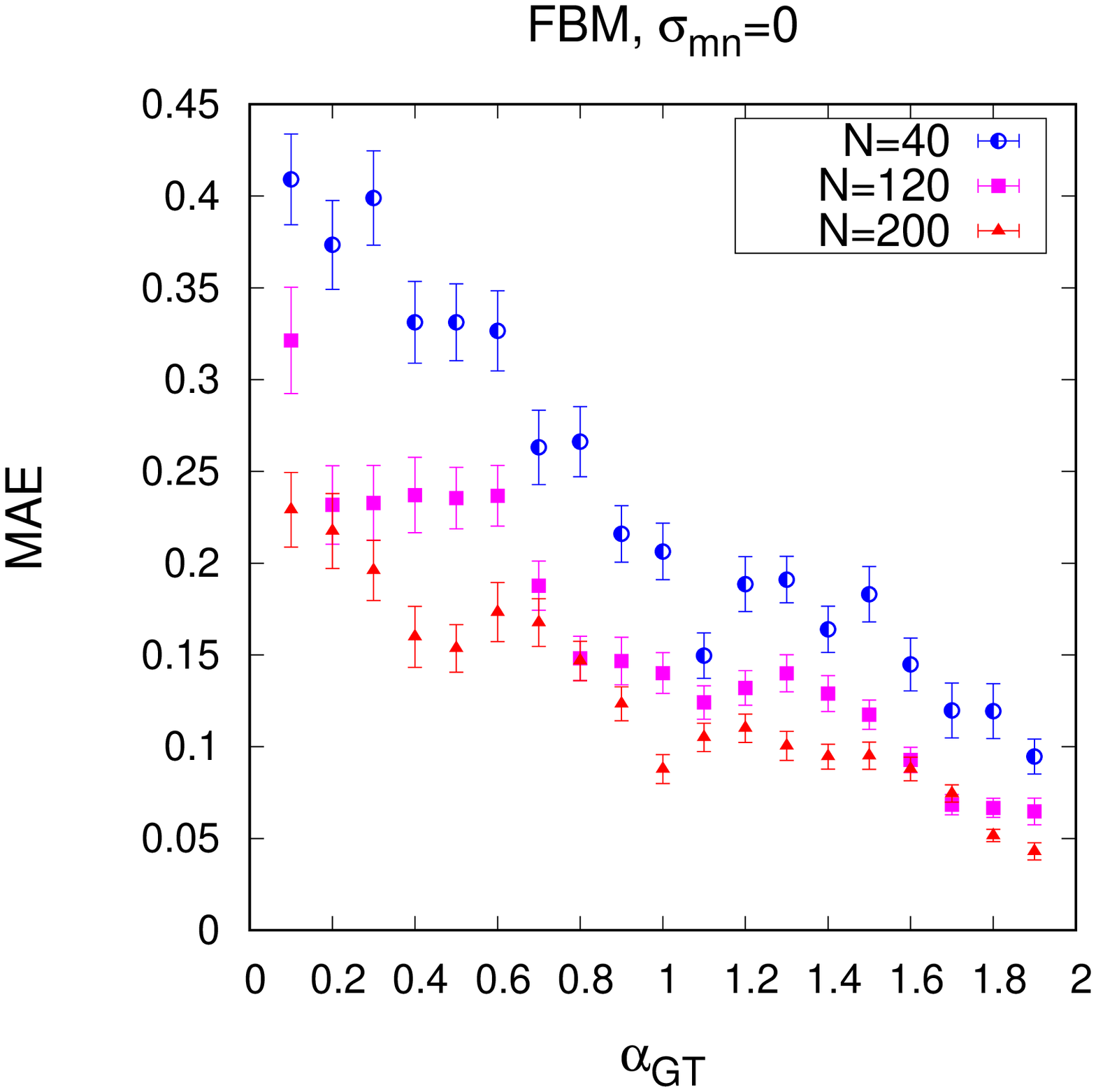} 
    \includegraphics[angle=270,scale=0.35]{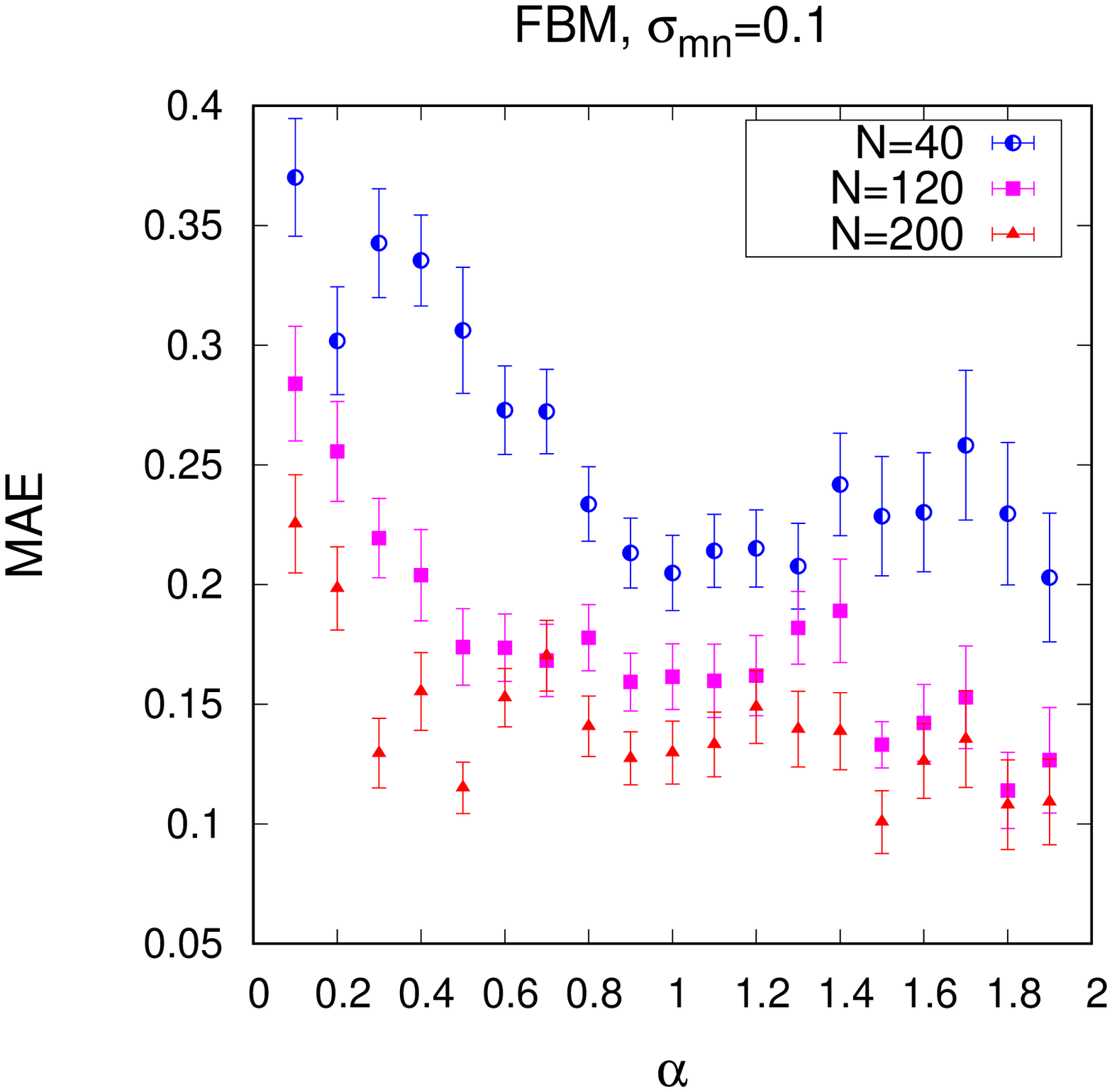}\\ 
    \includegraphics[angle=270,scale=0.35]{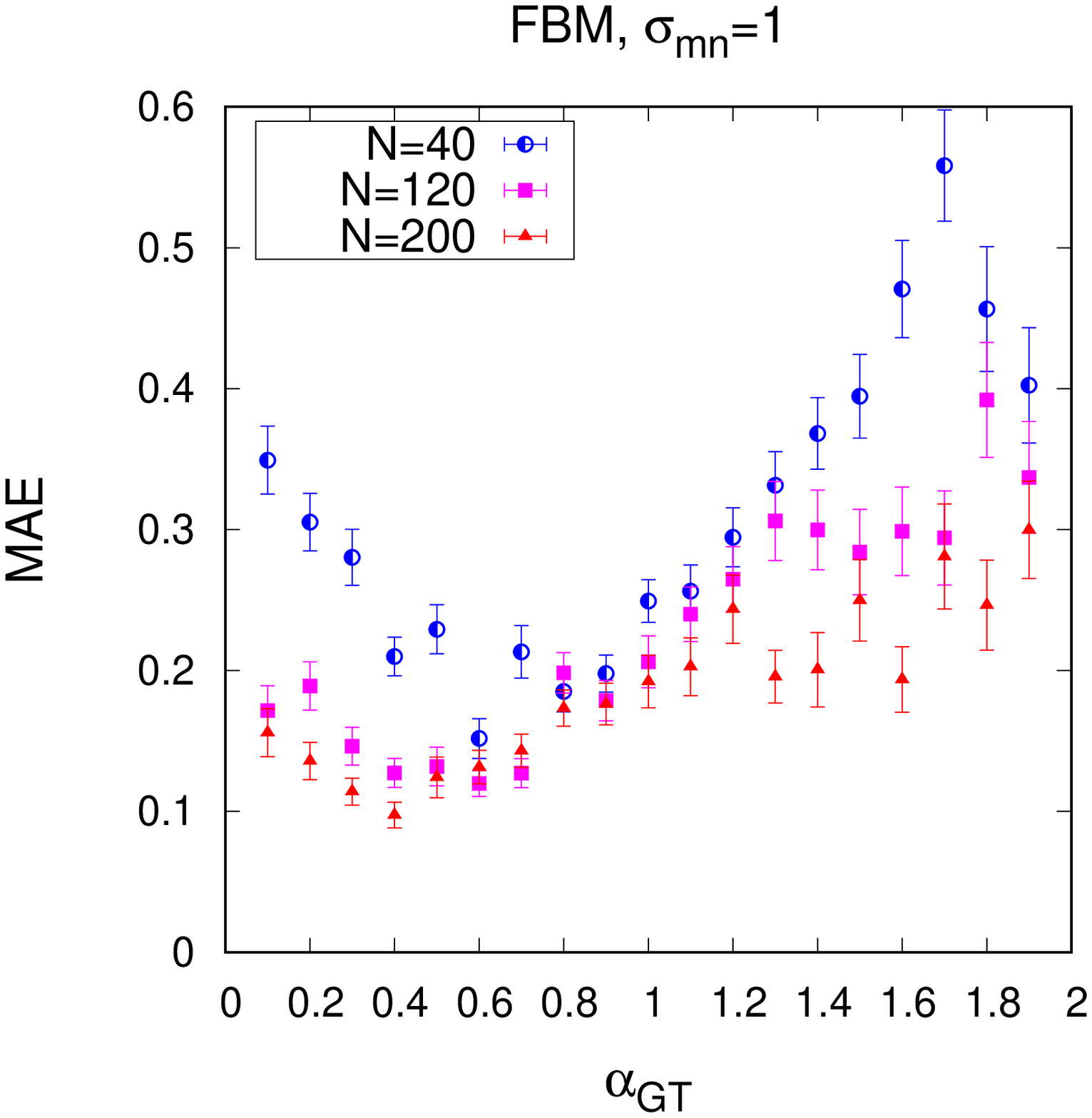} 
    \includegraphics[angle=270,scale=0.35]{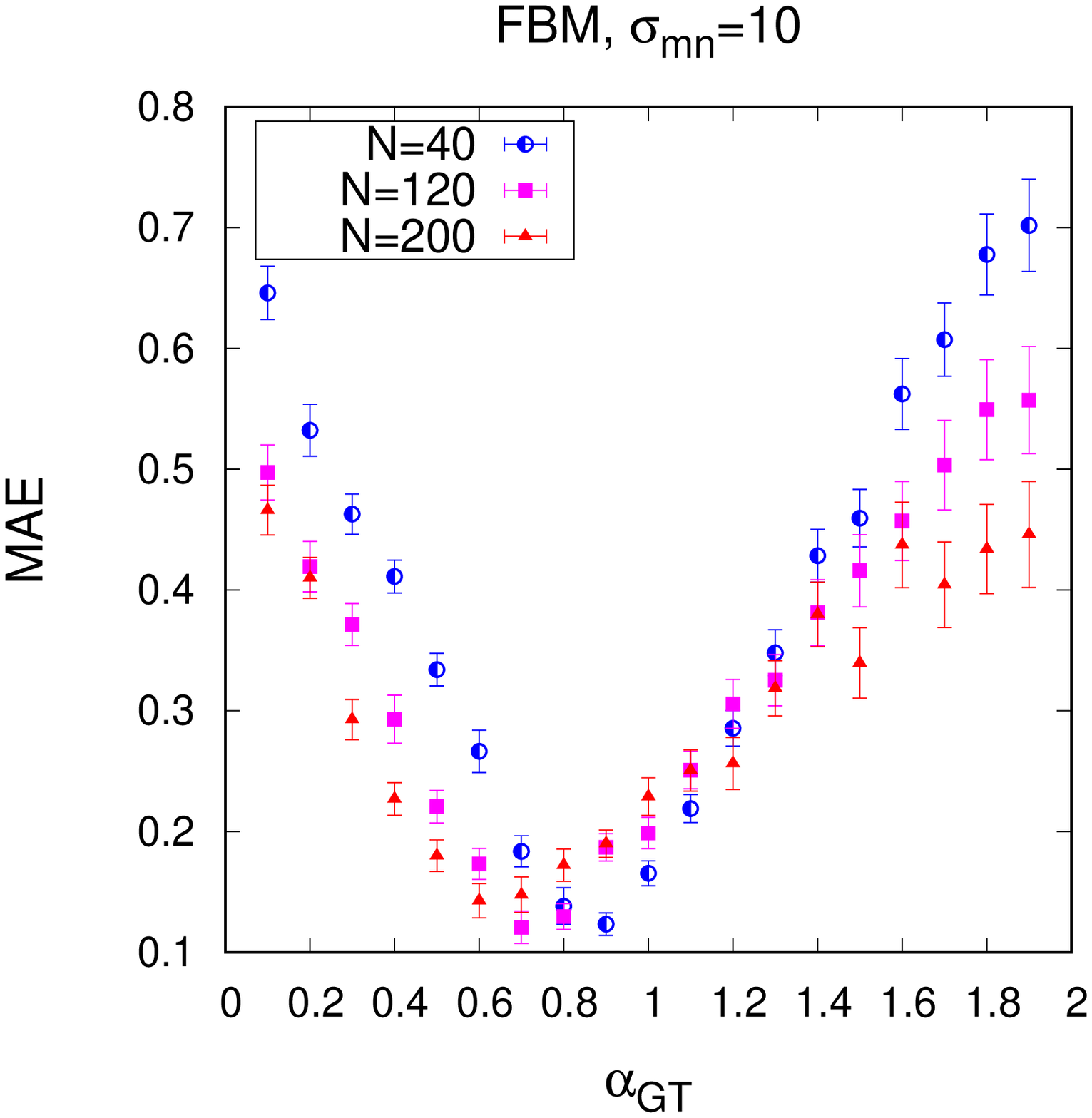} 
     \caption{MAE vs. $\alpha_{\rm GT}$ plot  with the estimated MAE obtained from the analysis of $\tilde{N}=100$ 1-dimensional FBM trajectories for each $\alpha_{\rm GT}$. Each sub-plot shows the results from the analysis of FBM trajectories generated with different noise-strengths $\sigma_{\rm mn}$. The error-bars are the standard error on the mean. This figure is equivalent to figure \ref{fig-mae-fbm-2d}, except that it is for 1-dimensional trajectories instead of 2-dimensional ones.}
    \label{fig-mae-fbm-1d}
\end{figure}

\begin{figure}
    \includegraphics[angle=270,scale=0.35]{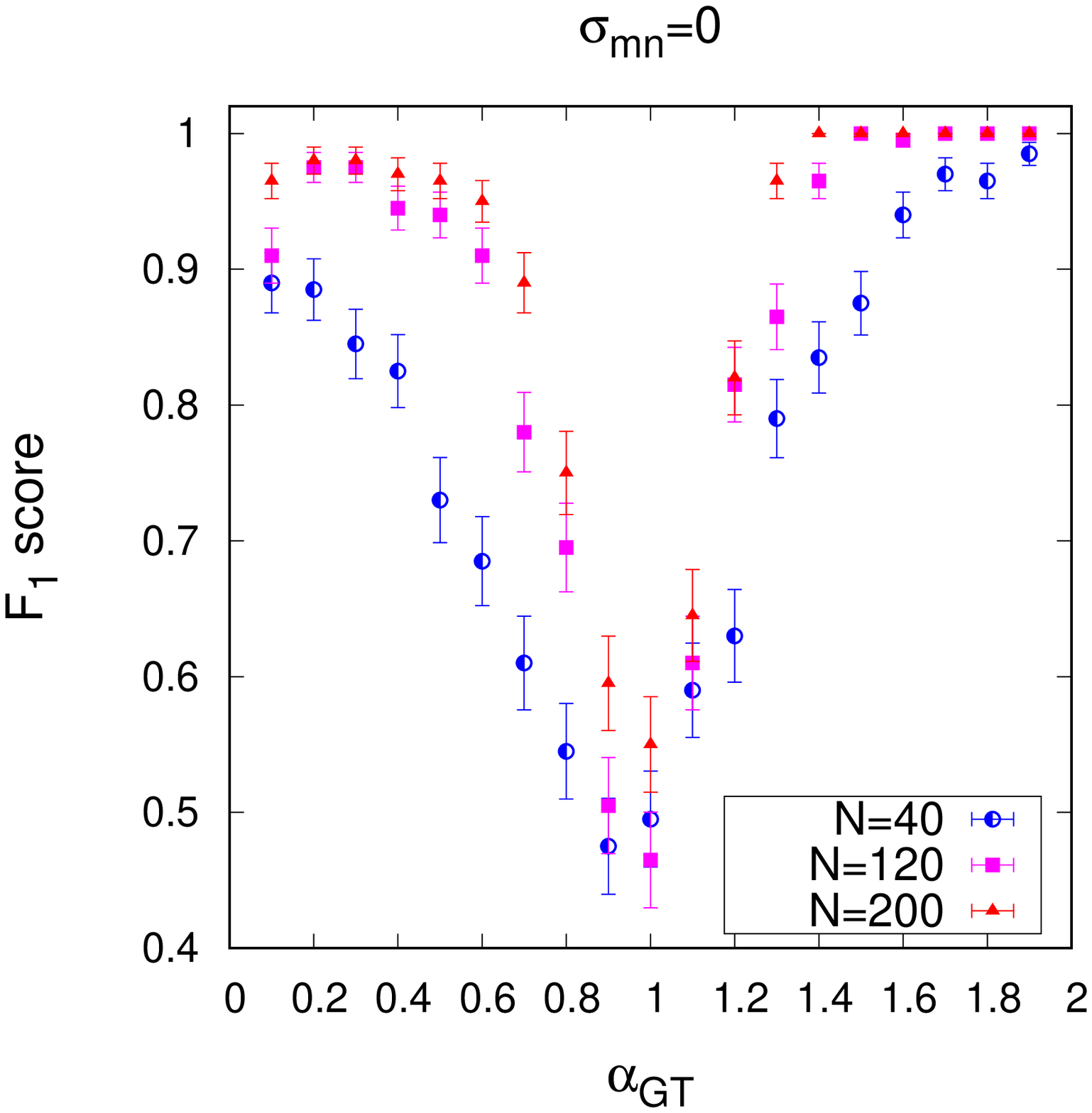}
    \includegraphics[angle=270,scale=0.35]{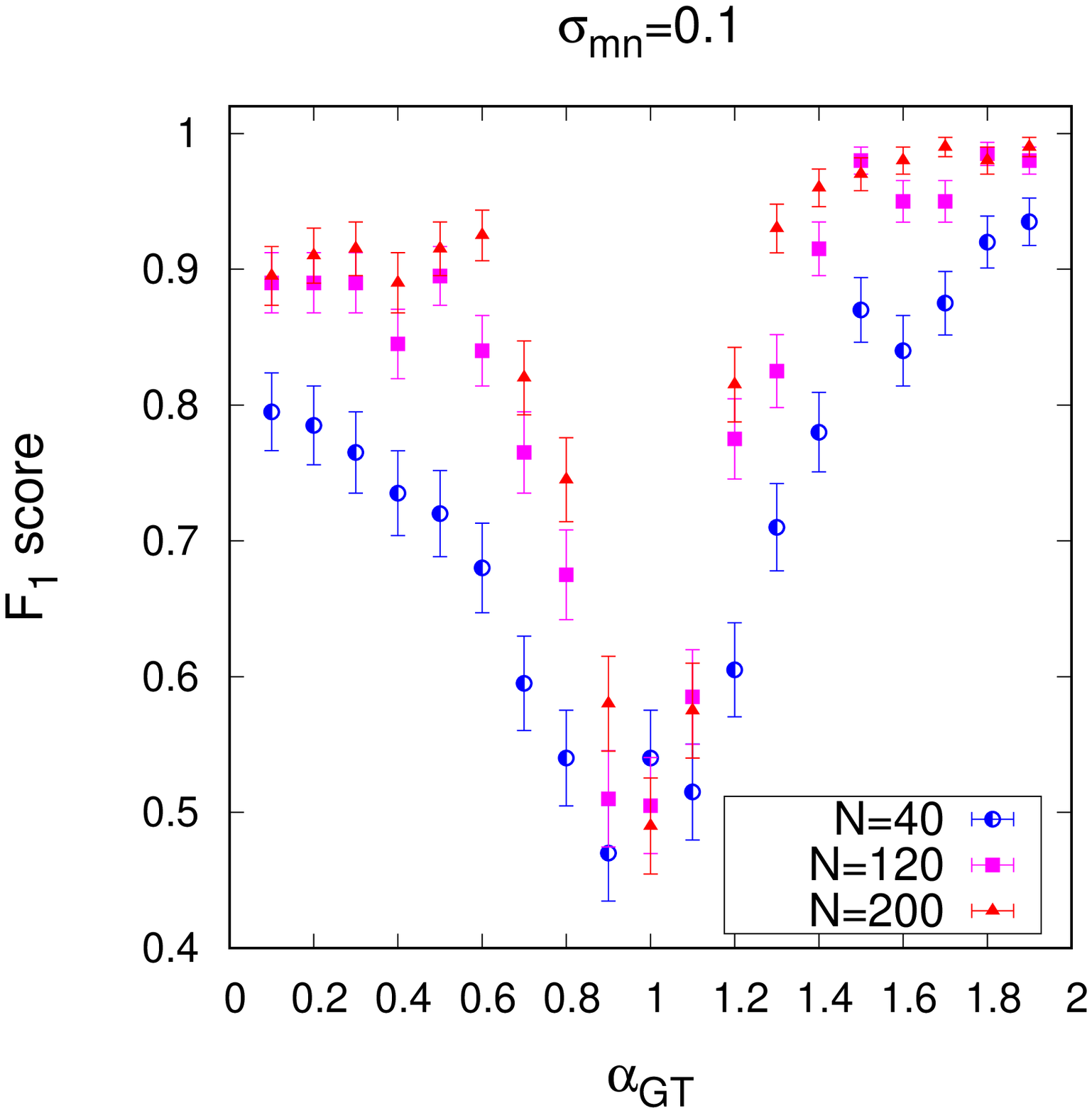}\\ 
    \includegraphics[angle=270,scale=0.35]{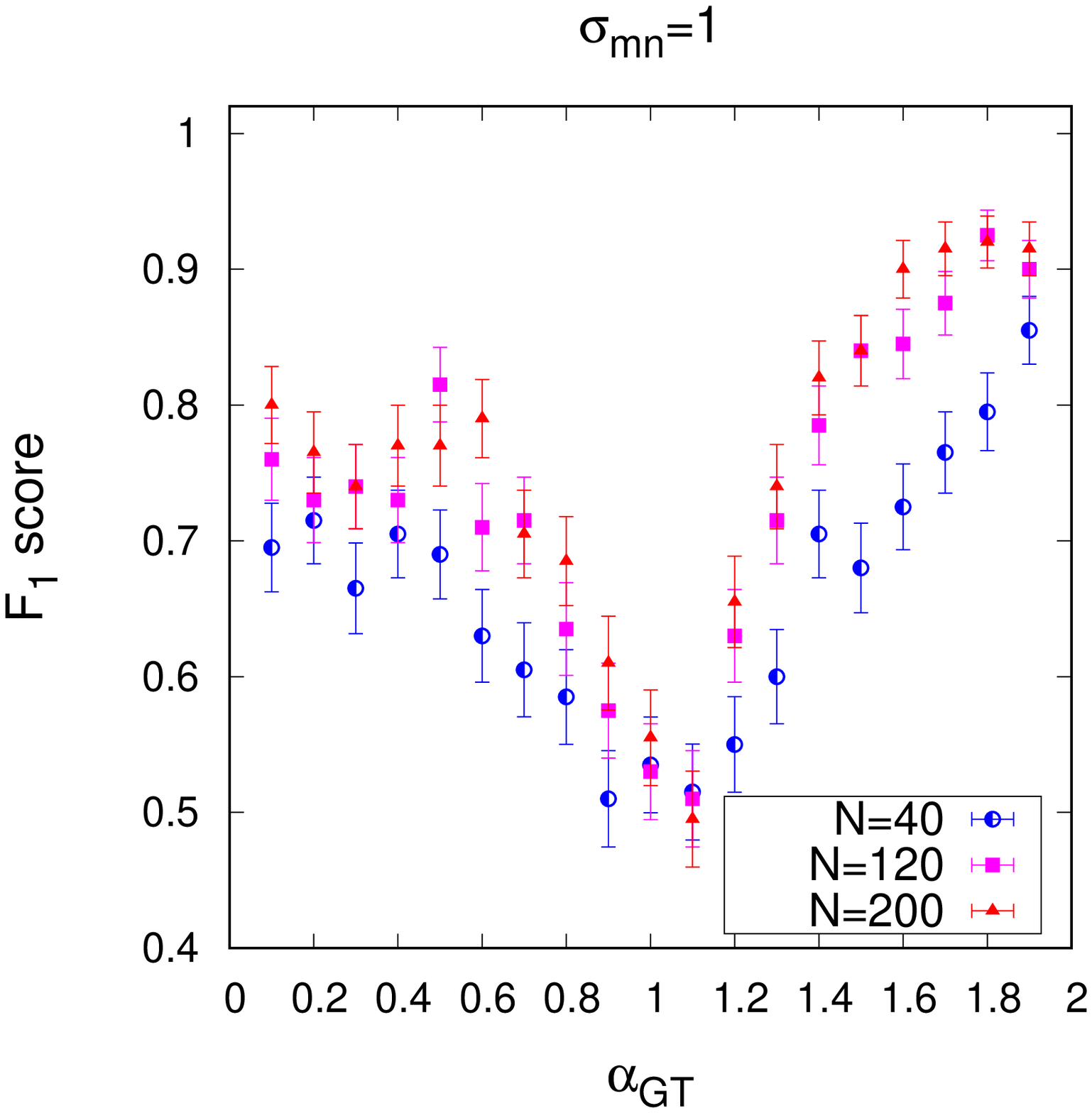} 
    \includegraphics[angle=270,scale=0.35]{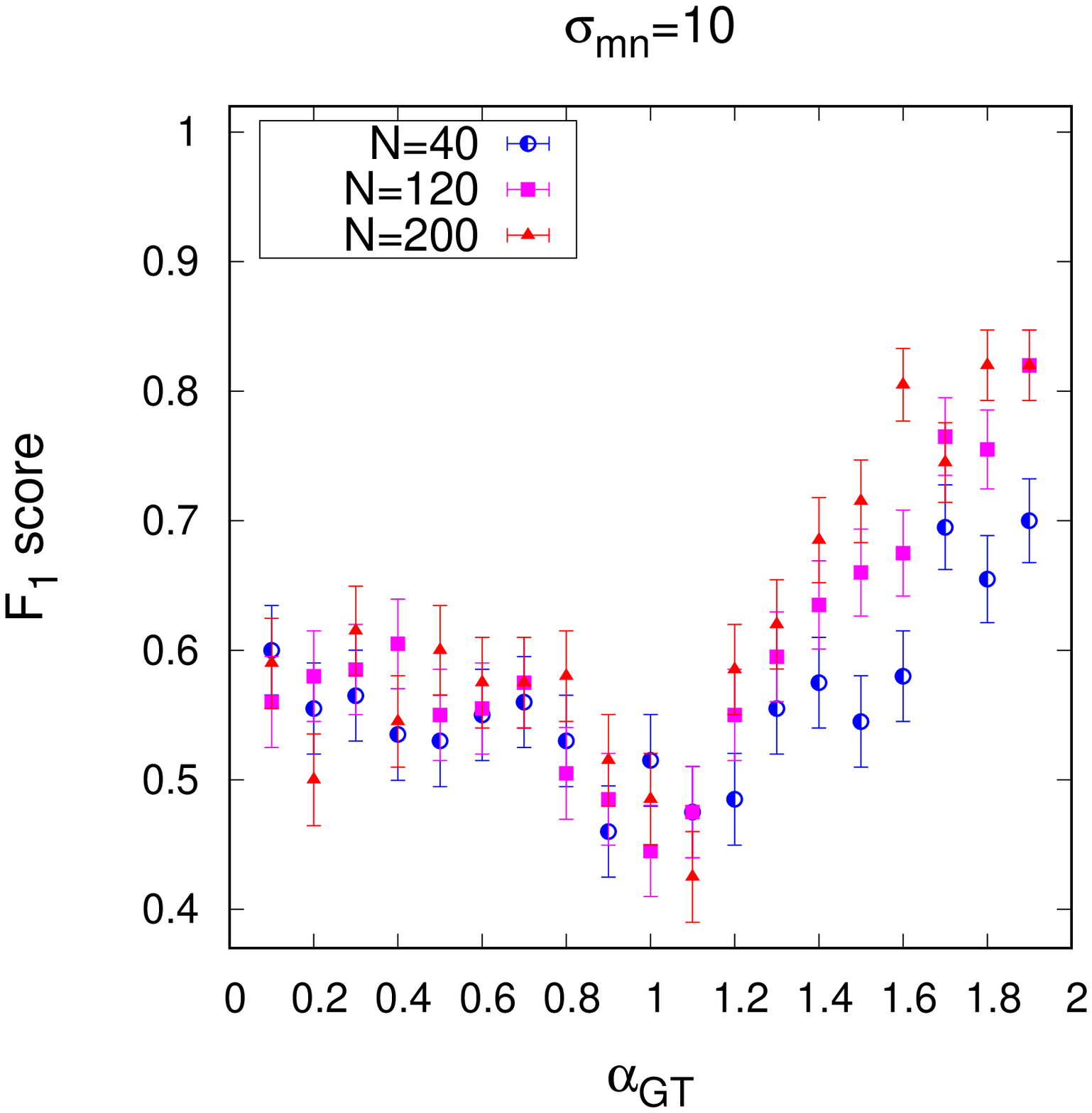}
   \caption{$F_1$ score vs. $\alpha_{\rm GT}$ plot from the analysis of 1-dimensional FBM  and SBM trajectories. 
    The $F_1$ scores at each $\alpha_{\rm GT}$ are from the analysis of $\tilde{N}=100$ SBM and $\tilde{N}=100$ FBM trajectories generated with that particular $\alpha_{\rm GT}$.
    The error-bars are the standard error on the mean for a Bernoulli random variable. This figure is equivalent to figure \ref{fig-f1-2d}, except that it is for 1-dimensional trajectories instead of 2-dimensional ones.}
    \label{fig-f1-1d}
\end{figure}

\clearpage

\section*{References}
\bibliographystyle{unsrt} 
\bibliography{sbm_inference}

\begin{thebibliography}{10}

\bibitem{brown1828}
R.~Brown.
\newblock A brief account of microscopical observations made in the months of
  june, july and august, 1827, on the particles contained in the pollen of
  plants; and on the general existence of active molecules in organic and
  inorganic bodies.
\newblock {\em Philos. Mag.}, 4:161, 1828.

\bibitem{einstein05}
A.~Einstein.
\newblock {\"U}ber die von der molekularkinetischen theorie der w{\"a}rme
  geforderte bewegung von in ruhenden fl{\"u}ssigkeiten suspendierten teilchen.
\newblock {\em Ann. Phys.}, 322:549, 1905.

\bibitem{perrin09}
J.~Perrin.
\newblock Movement brownien et r{\'e}alit{\'e} moleculaire (bronwian motion and
  molecular reality).
\newblock {\em Ann. Chim. Phys.}, 18:5, 1909.

\bibitem{nordlund14}
I.~Nordlund.
\newblock Eine neue bestimmung der avogadroschen konstante aus der brownschen
  bewegung kleiner, in wasser suspendierten quecksilberk{\"u}gelchen (a new
  determination of avogadro's constant from brown's motion of small mercury
  globules suspended in water).
\newblock {\em Z. Phys. Chem.}, 87:40, 1914.

\bibitem{metzler2014}
R.~Metzler, J.~H. Jeon, A.~G. Cherstvy, and E.~Barkai.
\newblock Anomalous diffusion models and their properties: non-stationarity{,}
  non-ergodicity{,} and ageing at the centenary of single particle tracking.
\newblock {\em Phys. Chem. Chem. Phys.}, 16:24128, 2014.

\bibitem{sokolov12}
I.~M. Sokolov.
\newblock Models of anomalous diffusion in crowded environments.
\newblock {\em Soft Matter}, 8:9043, 2012.

\bibitem{mackay03}
D.~J.~C. MacKay.
\newblock {\em Information Theory, Inference, and Learning Algorithms}.
\newblock Cambridge University Press, 2003.

\bibitem{sivia06}
D.~S. Sivia and J.~Skilling.
\newblock {\em Data Analysis: A Bayesian Tutorial}.
\newblock Oxford University Press, 2. edition, 2006.

\bibitem{robert07}
C.~P. Robert.
\newblock {\em The Bayesian Choice}.
\newblock Springer, 2. edition, 2007.

\bibitem{gelman13}
A.~Gelman \textit{et al.}
\newblock {\em Bayesian Data Analysis}.
\newblock CRC Press, 2. edition, 2013.

\bibitem{elliott16}
G.~Elliott and A.~Timmermann.
\newblock {\em Economic Forecasting}.
\newblock Princeton University Press, 2016.

\bibitem{andi21}
G.~Muñoz-Gil \textit{et al.}
\newblock Objective comparison of methods to decode anomalous diffusion.
\newblock {\em Nat. Comm.}, 12:6253, 2021.

\bibitem{krog17}
J.~Krog and M.~A. Lomholt.
\newblock Bayesian inference with information content model check for langevin
  equations.
\newblock {\em Phys. Rev. E}, 96:062106, 2017.

\bibitem{krog18}
J.~Krog, L.~H. Jacobsen, F.~W. Lund, D.~W{\"u}stner, and M.~A. Lomholt.
\newblock Bayesian model selection with fractional brownian motion.
\newblock {\em J. Stat. Mech.}, 2018:093501, 2018.

\bibitem{thapa2018}
S.~Thapa, M.~A. Lomholt, J.~Krog, A.~G. Cherstvy, and R.~Metzler.
\newblock Bayesian analysis of single-particle tracking data using the
  nested-sampling algorithm: maximum-likelihood model selection applied to
  stochastic-diffusivity data.
\newblock {\em Phys. Chem. Chem. Phys.}, 20:29018, 2018.

\bibitem{park2021}
S.~Park, S.~Thapa, Y.~Kim, M.~A. Lomholt, and J.~H. Jeon.
\newblock Bayesian inference of l\'evy walks via hidden markov models.
\newblock {\em J. Phys. A: Math. Theor.}, 54:484001, 2021.

\bibitem{lim02}
S.~C. Lim and S.~V. Muniandy.
\newblock Self-similar gaussian processes for modeling anomalous diffusion.
\newblock {\em Phys. Rev. E}, 66:021114, 2002.

\bibitem{saxton01}
M.~J. Saxton.
\newblock Anomalous subdiffusion in fluorescence photobleaching recovery: a
  monte carlo study.
\newblock {\em Biophys. J.}, 81:2226, 2001.

\bibitem{mandelbrot68}
B.~B. Mandelbrot and J.~W. Van~Ness.
\newblock Fractional brownian motions, fractional noises and applications.
\newblock {\em SIAM review}, 10(4):422, 1968.

\bibitem{weiss07}
G.~Guigas, C.~Kalla, and M.~Weiss.
\newblock The degree of macromolecular crowding in the cytoplasm and
  nucleoplasm of mammalian cells is conserved.
\newblock {\em FEBS Lett.}, 581:5094, 2007.

\bibitem{verkman98}
N.~Periasmy and A.~S. Verkman.
\newblock Analysis of fluorophore diffusion by continuous distributions of
  diffusion coefficients: Application to photobleaching measurements of
  multicomponent and anomalous diffusion.
\newblock {\em Biophys. J.}, 75:557, 1998.

\bibitem{berland08}
J.~Wu and M.~Berland.
\newblock Propagators and time-dependent diffusion coefficients for anomalous
  diffusion.
\newblock {\em Biophys. J.}, 95:2049, 2008.

\bibitem{hoyst06}
J.~Szymaski, A.~Patkowski, J.~Gapiski, A.~Wilk, and R.~Hoyst.
\newblock Movement of proteins in an environment crowded by surfactant
  micelles: anomalous versus normal diffusion.
\newblock {\em J. Phys. Chem. B.}, 110:7367, 2006.

\bibitem{doussal92}
P.~P. Mitra, P.~N. Sen, L.~M. Schwartz, and P.~Le Doussal.
\newblock Diffusion propagator as a probe of the structure of porous media.
\newblock {\em Phys. Rev. Lett.}, 68:3555, 1992.

\bibitem{boon13}
J.~F. Lutsko and J.~P. Boon.
\newblock Microscopic theory of anomalous diffusion based on particle
  interactions.
\newblock {\em Phys. Rev. E}, 88:022108, 2013.

\bibitem{hyung14}
J.~H. Jeon, A.~V. Chechkin, and R.~Metzler.
\newblock Scaled brownian motion: a paradoxical process with a time dependent
  diffusivity for the description of anomalous diffusion.
\newblock {\em Phys. Chem. Chem. Phys.}, 16:15811, 2014.

\bibitem{safdari2015}
H.~Safdari, A.~G. Cherstvy, A.~V. Chechkin, F.~Thiel, I.~M. Sokolov, and
  R.~Metzler.
\newblock Quantifying the non-ergodicity of scaled {B}rownian motion.
\newblock {\em J. Phys. A: Math. Theor.}, 48:375002, 2015.

\bibitem{cherstvy_sbm2015}
A.~G. Cherstvy and R.~Metzler.
\newblock Ergodicity breaking, ageing, and confinement in generalized diffusion
  processes with position and time dependent diffusivity.
\newblock {\em J. Stat. Mech.}, 2015:P05010, 2015.

\bibitem{bodrova2015}
A.~Bodrova, A.~V. Chechkin, A.~G. Cherstvy, and R.~Metzler.
\newblock Quantifying non-ergodic dynamics of force-free granular gases.
\newblock {\em Phys. Chem. Chem. Phys.}, 17:21791, 2015.

\bibitem{michalet10}
X.~Michalet.
\newblock Mean square displacement analysis of single-particle trajectories
  with localization error: Brownian motion in an isotropic medium.
\newblock {\em Phys. Rev. E}, 82:041914, 2010.

\bibitem{michalet12}
X.~Michalet and A.~J. Berglund.
\newblock Optimal diffusion coefficient estimation in single-particle tracking.
\newblock {\em Phys. Rev. E}, 85:061916, 2012.

\bibitem{mandelbrot82}
B.~B. Mandelbrot.
\newblock {\em The fractal geometry of nature}.
\newblock W. H. Freeman, New York, 1982.

\bibitem{cherstvy2019}
A.~G. Cherstvy, S.~Thapa, C.~E. Wagner, and R.~Metzler.
\newblock Non-gaussian{,} non-ergodic{,} and non-fickian diffusion of tracers
  in mucin hydrogels.
\newblock {\em Soft Matter}, 15:2526, 2019.

\bibitem{jeon2011}
J.~H. Jeon, V.~Tejedor, S.~Burov, E.~Barkai, C.~Selhuber-Unkel,
  K.~Berg-S\o{}rensen, L.~Oddershede, and R.~Metzler.
\newblock In vivo anomalous diffusion and weak ergodicity breaking of lipid
  granules.
\newblock {\em Phys. Rev. Lett.}, 106:048103, Jan 2011.

\bibitem{weiss09}
J.~Szymanski and M.~Weiss.
\newblock Elucidating the origin of anomalous diffusion in crowded fluids.
\newblock {\em Phys. Rev. Lett.}, 103:038102, 2009.

\bibitem{weron09}
M.~Magdziarz, A.~Weron, K.~Burnecki, and J.~Klafter.
\newblock Fractional brownian motion versus the continuous-time random walk: A
  simple test for subdiffusive dynamics.
\newblock {\em Phys. Rev. Lett.}, 103:180602, 2009.

\bibitem{klafter10}
M.~Magdziarz and J.~Klafter.
\newblock Detecting origins of subdiffusion: p-variation test for confined
  systems.
\newblock {\em Phys. Rev. E}, 82:011129, 2010.

\bibitem{burn10}
K.~Burnecki and J.~Klafter.
\newblock Fractional l{\'e}vy stable motion can model subdiffusive dynamics.
\newblock {\em Phys. Rev. E}, 82:021130, 2010.

\bibitem{weber10}
S.~C. Weber, A.~J. Spakowitz, and J.~A. Theriot.
\newblock Bacterial chromosomal loci move subdiffusively through a viscoelastic
  cytoplasm.
\newblock {\em Phys. Rev. Lett.}, 104:238102, 2010.

\bibitem{korabel20}
D.~Han, N.~Korabel, R.~Chen, M.~Johnston, A.~Gavrilova, V.~J. Allan,
  S.~Fedotov, and T.~A. Waigh.
\newblock Deciphering anomalous heterogeneous intracellular transport with
  neural networks.
\newblock {\em eLife}, 9:e52224, 2020.

\bibitem{weiss-ent-21}
K.~Speckner and M.~Weiss.
\newblock Single-particle tracking reveals anti-persistent subdiffusion in cell
  extracts.
\newblock {\em Entropy}, 23:892, 2021.

\bibitem{weiss-njp-21}
R.~Benelli and M.~Weiss.
\newblock From sub- to superdiffusion: fractional brownian motion of
  membraneless organelles in early c. elegans embryos.
\newblock {\em New J. Phys.}, 23:063072, 2021.

\bibitem{deng09}
Weihua Deng and Eli Barkai.
\newblock Ergodic properties of fractional brownian-langevin motion.
\newblock {\em Phys. Rev. E}, 79:011112, 2009.

\bibitem{jeon2012}
J.~H. Jeon and R.~Metzler.
\newblock Inequivalence of time and ensemble averages in ergodic systems:
  exponential versus power-law relaxation in confinement.
\newblock {\em Phys. Rev. E}, 85:021147, 2012.

\bibitem{jeon2013}
J.~H. Jeon, N.~Leijnse, L.~B. Oddershede, and R.~Metzler.
\newblock Anomalous diffusion and power-law relaxation of the time averaged
  mean squared displacement in worm-like micellar solutions.
\newblock {\em New J. Phys.}, 15:045011, 2013.

\bibitem{manzo2015}
C.~Manzo and M.~F. Garcia-Parajo.
\newblock A review of progress in single particle tracking: from methods to
  biophysical insights.
\newblock {\em Rep. Prog. Phys.}, 78:124601, 2015.

\bibitem{lene2017}
K.~Norregaard, R.~Metzler, C.~M. Ritter, K.~Berg-Sørensen, , and L.~B.
  Oddershede.
\newblock Manipulation and motion of organelles and single molecules in living
  cells.
\newblock {\em Chem. Rev.}, 117:4342, 2017.

\bibitem{skilling04}
J.~Skilling.
\newblock Nested sampling.
\newblock {\em AIP Conf. Proc.}, 735:395, 2004.

\bibitem{skilling06}
J.~Skilling.
\newblock Nested sampling for general bayesian computation.
\newblock {\em Bayesian Analysis}, 1:833--859, 2006.

\bibitem{nishimura20}
A.~Nishimura, D.~B. Dunson, and J.~Lu.
\newblock Discontinuous hamiltonian monte carlo for discrete parameters and
  discontinuous likelihoods.
\newblock {\em Biometrika}, 107:365, 2020.

\bibitem{sbm_git}
Our code is available at: \verb"https://github.com/Samudrajit11/SBM".

\bibitem{krapf11}
A.~V. Weigel, B.~Simon, M.~M. Tamkun, and D.~Krapf.
\newblock Ergodic and nonergodic processes coexist in the plasma membrane as
  observed by single-molecule tracking.
\newblock {\em Proc. Natl. Acad. Sci. U.S.A.}, 108:6438, 2011.

\bibitem{weitz04}
I.~Y. Wong, M.~L. Gardel, D.~R. Reichman, E.~R. Weeks, M.~T. Valentine, A.~R.
  Bausch, and D.~A. Weitz.
\newblock Anomalous diffusion probes microstructure dynamics of entangled
  f-actin networks.
\newblock {\em Phys. Rev. Lett.}, 92:178101, 2004.

\bibitem{chaikin11}
Q.~Xu, L.~Feng, R.~Sha, N.~C. Seeman, and P.~M. Chaikin.
\newblock Subdiffusion of a sticky particle on a surface.
\newblock {\em Phys. Rev. Lett.}, 106:228102, 2011.

\bibitem{swinney93}
T.~H. Solomon, E.~R. Weeks, and H.~L. Swinney.
\newblock Observation of anomalous diffusion and lévy flights in a
  two-dimensional rotating flow.
\newblock {\em Phys. Rev. Lett.}, 71:3975, 1993.

\bibitem{scher1975}
H.~Scher and E.~W. Montroll.
\newblock Anomalous transit-time dispersion in amorphous solids.
\newblock {\em Phys. Rev. B}, 12:2455, 1975.

\bibitem{montroll1969}
E.~W. Montroll.
\newblock Random walks on lattices. iii. calculation of first‐passage times
  with application to exciton trapping on photosynthetic units.
\newblock {\em Journal of Mathematical Physics}, 10:753, 1969.

\bibitem{hans07}
D.~Kleinhans and R.~Friedrich.
\newblock Continuous-time random walks: Simulation of continuous trajectories.
\newblock {\em Phys. Rev. E}, 76:061102, 2007.

\bibitem{thiel2014}
F.~Thiel and I.~M. Sokolov.
\newblock Scaled brownian motion as a mean-field model for continuous-time
  random walks.
\newblock {\em Phys. Rev. E}, 89:012115, 2014.

\bibitem{khosravi11}
A.~Khosravi, S.~Nahavandi, D.~Creighton, and A.~F. Atiya.
\newblock Comprehensive review of neural network-based prediction intervals and
  new advances.
\newblock {\em IEEE Trans. Neural Netw.}, 22(9):1341, 2011.

\end{thebibliography}

\end{document}